\documentclass[final,3p,times,onecolumn,numafflabel,super]{elsarticle}
\biboptions{super,sort&compress}

\usepackage[utf8]{inputenc}
\usepackage{textcomp}
\usepackage{lmodern,textcomp} 
\usepackage{subfig}
\usepackage[-]{callouts}
\usepackage{tikz}
\usetikzlibrary{decorations.pathreplacing,calligraphy}

\graphicspath{{graphics/}}

\DeclareGraphicsExtensions{.pdf,.jpeg,.png}

\usepackage{epsfig}
\usepackage{lipsum}

\usepackage{amsmath}
\usepackage{amsfonts}
\usepackage{amssymb}
\usepackage{float}
\usepackage[hyphens]{url}

\usepackage[normalem]{ulem}

\usepackage{booktabs}
\usepackage{tabularx}
\usepackage{threeparttable}
\usepackage{siunitx}
\usepackage[version=4]{mhchem}
\usepackage[normalem]{ulem}

\usepackage{url}
\usepackage[colorlinks=true, citecolor=blue, linkcolor=blue, filecolor=blue,urlcolor=blue,breaklinks=true]{hyperref}
\usepackage{enumitem}

\usepackage[gen]{eurosym}
\usepackage{multirow}

\usepackage{threeparttable}
\usepackage{makecell}

\usepackage{color, colortbl}

\def\co{CO${}_2$}

\makeatletter
\newcommand{\vast}{\bBigg@{3}}
\newcommand{\Vast}{\bBigg@{4}}
\makeatother

\usepackage{tikz}
\usepackage{adjustbox}

\usepackage{fixltx2e}
\usepackage{pdflscape}
\usepackage{changepage}

\journal{}

\begin{document}
\begin{frontmatter}

\title{E-biofuels reduce the cost of achieving emissions targets in hard-to-electrify sectors}

\author[chalmers]{Y.~Zhang\corref{cor1}}
\ead{yunlong.zhang@chalmers.se}
\author[rise,chalmers]{M.~Millinger\corref{cor1}}
\author[chalmers]{F.~Hedenus}
\author[rise]{K.~Pettersson}
\author[tuberlin]{T.~Brown}

\cortext[cor1]{Corresponding author}
\address[chalmers]{Department of Environmental and Energy Sciences, Chalmers University of Technology, 41296 Göteborg, Sweden}
\address[rise]{Built Environment: System Transition: Energy Systems Analysis, RISE Research Institutes of Sweden, Göteborg, Sweden}
\address[tuberlin]{Department of Digital Transformation in Energy Systems, Technische Universität Berlin, Berlin, Germany}

\begin{abstract}
Renewable liquid fuels are essential for achieving emissions targets for hard-to-electrify sectors such as aviation and shipping. While biofuels and synthetic e-fuels have been well-studied, e-biofuels, produced by adding renewable hydrogen to biomass conversion to better utilise the biogenic carbon, remain understudied and lack a clear role in EU fuel regulations. In this paper, using a sector-coupled European energy system model, we find that e-biofuels are cost-effective to meet stringent emissions targets if biomass availability is limited and fossil fuels are ineligible, either due to limited carbon sequestration capacity or to high renewable fuel mandates. By directly increasing utilisation of biogenic carbon instead of synthesising fuels based on captured \co, there are savings from fuel production and carbon capture that reduce total system costs by up to 2.7\% and liquid fuel costs by more than 10\%. Our results highlight the role of e-biofuels as a potential hedge against uncertainty in biomass, hydrogen, and carbon storage availability, as well as evolving policy implementation. 
\end{abstract}

\begin{keyword}
renewable fuels \sep sector-coupling \sep CCU \sep BECCS \sep biofuels \sep sustainable aviation fuel \sep SAF \sep e-fuels \sep e-biofuels
\end{keyword}

\end{frontmatter}

Achieving climate neutrality of hard-to-electrify sectors such as aviation, shipping and chemicals is one of the largest challenges in the energy system \cite{Millinger2025,Bachorz2025}. These sectors rely on energy-dense carbonaceous fuels, for which direct electrification is technologically limited or economically prohibitive. Reflecting this challenge, the European Union (EU) has set a target of 70\% sustainable aviation fuel (SAF) at European airports by 2050, half of which is to be synthetic fuels from renewable hydrogen and captured carbon \cite{EU2023_aviation}. In the maritime sectors, where more fuel alternatives exist and larger electrification shares are feasible, an 80\% reduction in greenhouse gas intensity is targeted by 2050 \cite{EU2023_maritime}. Achieving these targets requires the large-scale adoption of novel fuel conversion pathways.

Three fuel pathways are typically considered to meet emissions targets in these sectors \cite{Mignone2024}. Firstly, advanced biofuels produced from biomass residues offer a way to avoid relying on energy crops, which are dominant today and connected to concerns of land use change and food competition. Forest residues offer the largest potentials, but due to a carbon-to-hydrogen mismatch of biomass compared to the produced fuels, only around 30\% of biogenic carbon ends up in liquid fuels for thermochemical biofuel pathways such as Fischer–Tropsch synthesis, with the rest being separated as \co \ from the syngas \cite{millinger2022biofuel}. The second option is using this excess \co, or captured \co \ from other sources such as heat and power plant flue gases, to produce e-fuels by combining it with hydrogen, which requires capital- and energy-intensive carbon capture, hydrogen production, and conversion infrastructure. This option enhances usage of precious biogenic carbon and often emerges in models as an important option for achieving emissions targets, especially when aiming for high renewable fuel shares or if carbon sequestration capacity is limited \cite{millinger2022biofuel,Millinger2025,schreyer2025net}. Finally, next to biofuels and e-fuels, a third option is the continued use of fossil fuels combined with negative emissions elsewhere in the energy system, through bioenergy with carbon capture and storage (BECCS) or direct air capture and storage (DACCS) \cite{schreyer2025net,Mignone2024,Millinger2025}.

In both research studies and policy, an often overlooked fourth option to produce carbonaceous fuels and chemicals is to balance out the carbon-to-hydrogen mismatch in the biofuel process by adding hydrogen, to be able to directly utilise the biogenic carbon in a partially oxidised state as carbon monoxide without the need for capturing and transporting it to a dedicated conversion facility \cite{Mesfun2023}. Such fuels are termed electrobiofuels \cite{Millinger2025}, bioelectrofuels \cite{furusjo2022bioelectrofuels}, PBtL (Power-Biomass-to-Liquids) \cite{de2026biofuels} and co-processing \cite{EC_HydrogenDelegatedActs_QA_2024}, from here on denoted e-biofuels.

\begin{figure}[h]
    \centering
    \includegraphics[width=1\linewidth]{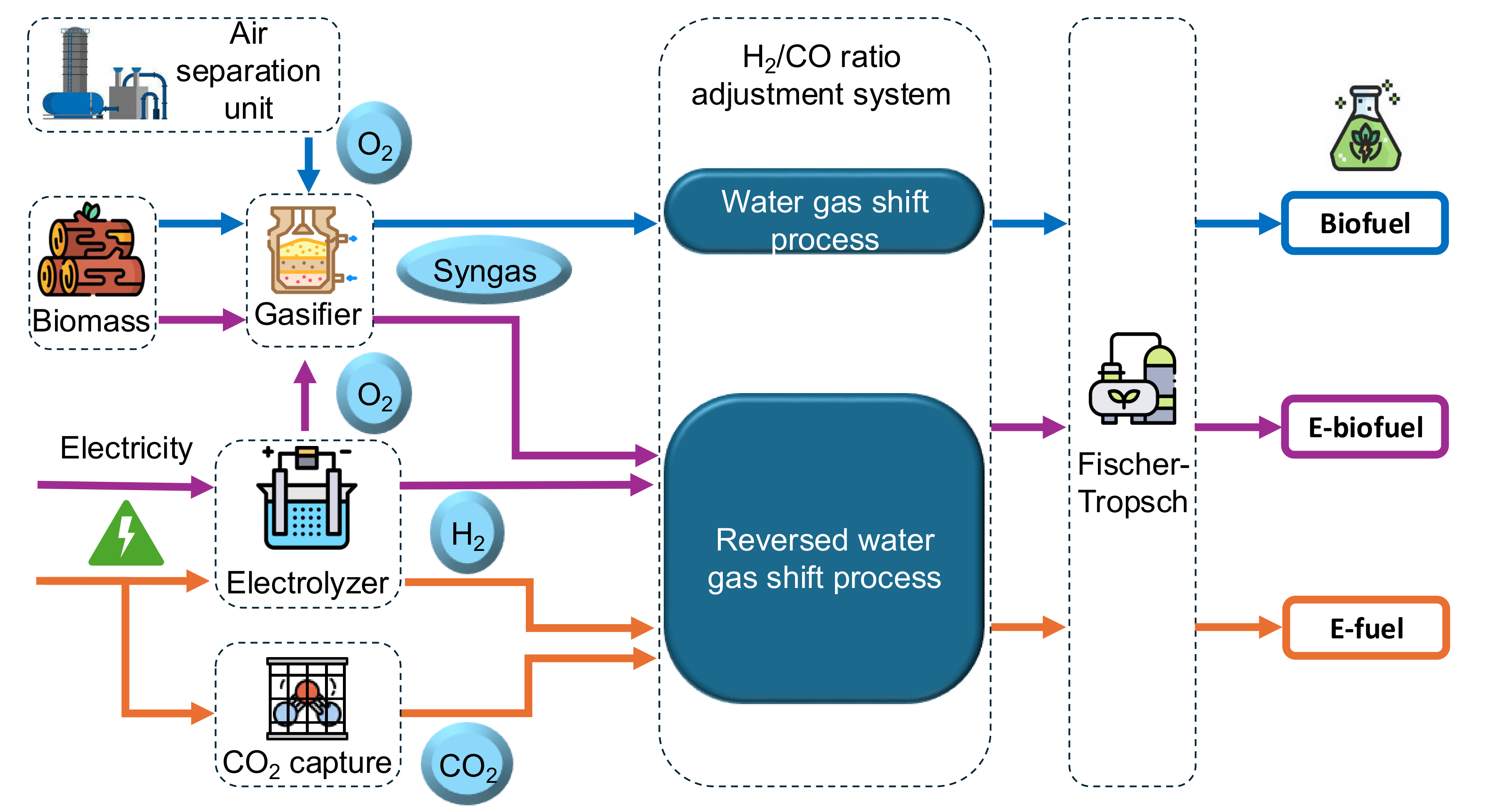}
    \caption{Schematic overview of three biomass-based renewable liquid fuel production pathways considered in this study: (i) conventional biofuels from biomass without hydrogen addition, (ii) e-fuels produced from renewable hydrogen and non-fossil \co\ sources, and (iii) e-biofuels, which integrate renewable hydrogen into biomass conversion to enhance carbon utilisation.}
    \label{fig:system_sketch}
\end{figure}

As illustrated in Fig.~\ref{fig:system_sketch}, e-biofuels integrate biomass gasification to syngas (a mixture of H$_2$ and CO) with additional hydrogen input, enabling both the utilisation of carbon required for conventional biofuels conversion and the further hydrogenation of excess biogenic carbon within the same system. This integrated hydrogenation concept is not limited to Fischer–Tropsch biomass-to-liquid (BtL) routes, but can similarly be applied to other syngas-based processes, such as biomass-to-methanol or biomass-to-SNG (synthetic natural gas). Moreover, biogas can serve as an alternative carbon-rich feedstock for e-biofuels: although the methane fraction still needs to be reformed to syngas \cite{bube2024power}, hydrogen-assisted adjustment of the H$_2$/CO ratio enables direct methanol synthesis without the need for biomass gasification. Process-level studies have consistently demonstrated the technical feasibility of these systems and shown that coupling biomass or biogas conversion with hydrogen boosting can substantially improve carbon efficiency and energy efficiency for biomass-to-liquid \cite{hillestad2018improving,ostadi2019boosting}, biomass-to-methanol \cite{anetjarvi2023benefits}, biomass-to-SNG \cite{katla2024synthetic} and biogas-to-methanol \cite{park2025bio} pathways. Near-term techno-economic assessments generally identify biofuels as the lowest-cost renewable fuel option, followed by e-biofuels and, subsequently, e-fuels \cite{Korberg2021b,furusjo2022bioelectrofuels,grahn2022review}. However, the economic performance of e-biofuels remains highly sensitive to hydrogen and carbon costs within the system framework. 

Currently, the scale-up of all four pathways faces challenges, as it relies on technologies unproven at scale, resources with uncertain availability subject to competition, and the establishment of novel value chains. Biomass availability is subject to sustainability risk and policy uncertainty; the availability of hydrogen for both e-fuels and e-biofuels is dependent on the scale-up of carbon-free electricity and electrolyzers \cite{Odenweller2022}; the alternative non-fossil carbon source from biomass or DAC is costly and require substantial carbon-free energy inputs, thereby also relying heavily on rapid expansion of renewable electricity such as wind and solar power, whose large-scale deployment itself faces growth constraints \cite{Cherp2021}. The possible continued use of fossil fuels in a net-zero energy system requires compensation by permanent removals (like-for-like principle\cite{Allen2022}), for which only technical CDR with geological sequestration (ie. geological storage) qualifies according to EU regulation \cite{EU2024_3012}, implying reliance on BECCS or DACCS. However, both carbon capture \cite{Kazlou2024} and carbon sequestration \cite{Grant2022,Gidden2025} are subject to concerns of scale-up speeds and available potentials.

Existing work on e-biofuels has largely focused on process-level simulations or isolated techno-economic analyses, while a system-wide assessment in a net-zero context remains limited. Although e-biofuels have been included in a few energy system modelling analyses \cite{glaum2025minMeOH,Millinger2025}, we still lack a comprehensive understanding of the cost-efficency of different liquid fuel options within integrated energy systems. Further, the current EU fuel policy does not explicitly recognise e-biofuels and, by design, tends to favour e-fuels \cite{Pettersson_RISE_REDFuels_2025}.

In this work, we provide a system-integrated assessment of the role of e-biofuels in supplying hard-to-electrify sectors within a net-zero European energy system. We show that e-biofuels become cost-effective under limited biomass and carbon sequestration availability, as the direct hydrogenation of excess biogenic carbon reduces system-wide fuel conversion and carbon capture capacity. E-biofuels reduce the whole energy system cost of achieving renewable fuel mandates in aviation and shipping by up to 3.1\%.

\clearpage
\section*{E-biofuels are cost-effective when phasing out fossil fuels}
\begin{figure}[h]
    \centering
    \includegraphics[width=1\linewidth]{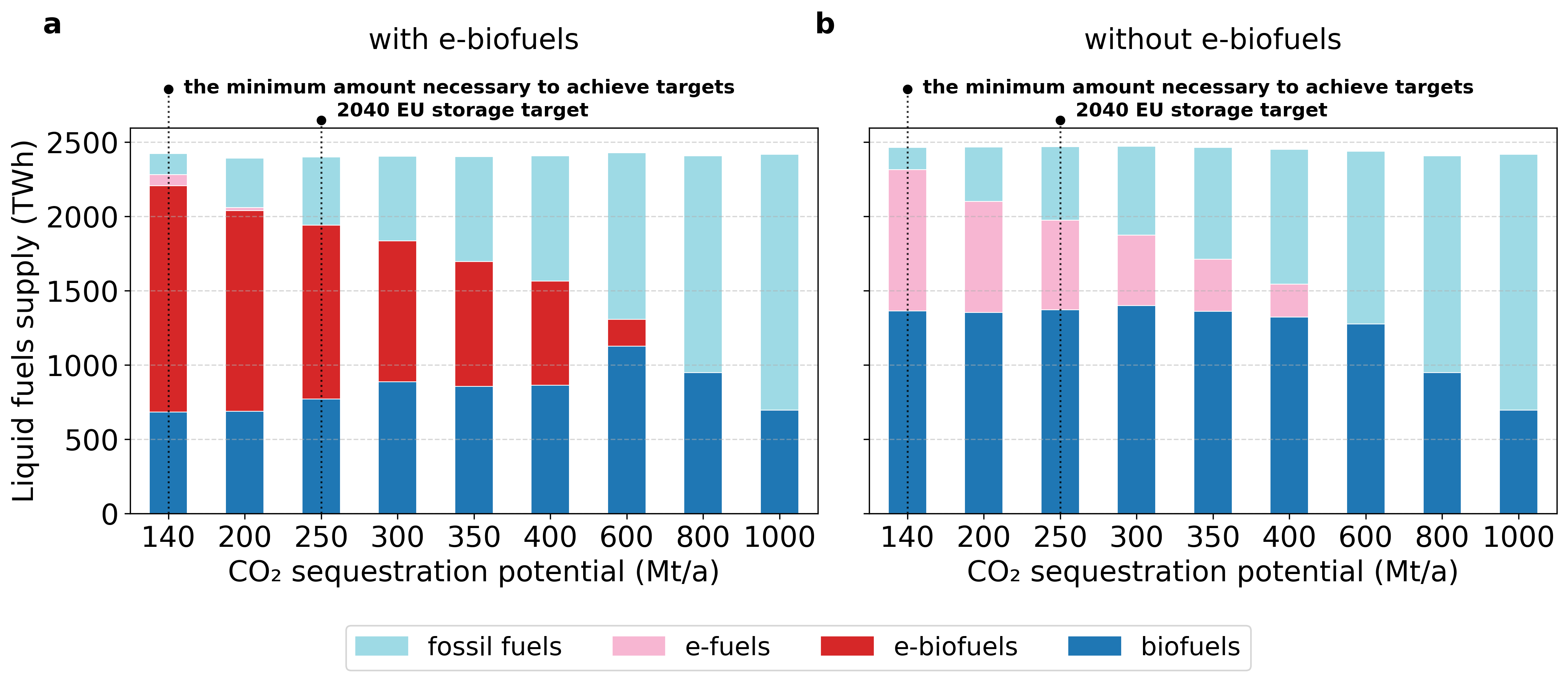}
    \caption{Liquid fuel mix in the European energy system under different annual carbon sequestration (CS). Panels (a) and (b) show the deployment of the four main categories of liquid fuels under the baseline scenario (without EU renewable fuels mandate constraint), \textit{with} and \textit{without} e-biofuels, respectively. The specific explanation of each fuel category can be found in the (Supplementary Table~\ref{tab:S1_liquid_fuel_definitions}). Total liquid fuel supply varies slightly across scenarios due to the distinct conversion efficiencies of methanol and oil pathways in meeting a fixed aviation kerosene demand.}
    \label{fig:fuelmix_CSsweep}
\end{figure}

\begin{figure}[h]
    \centering
    \includegraphics[width=1\linewidth]{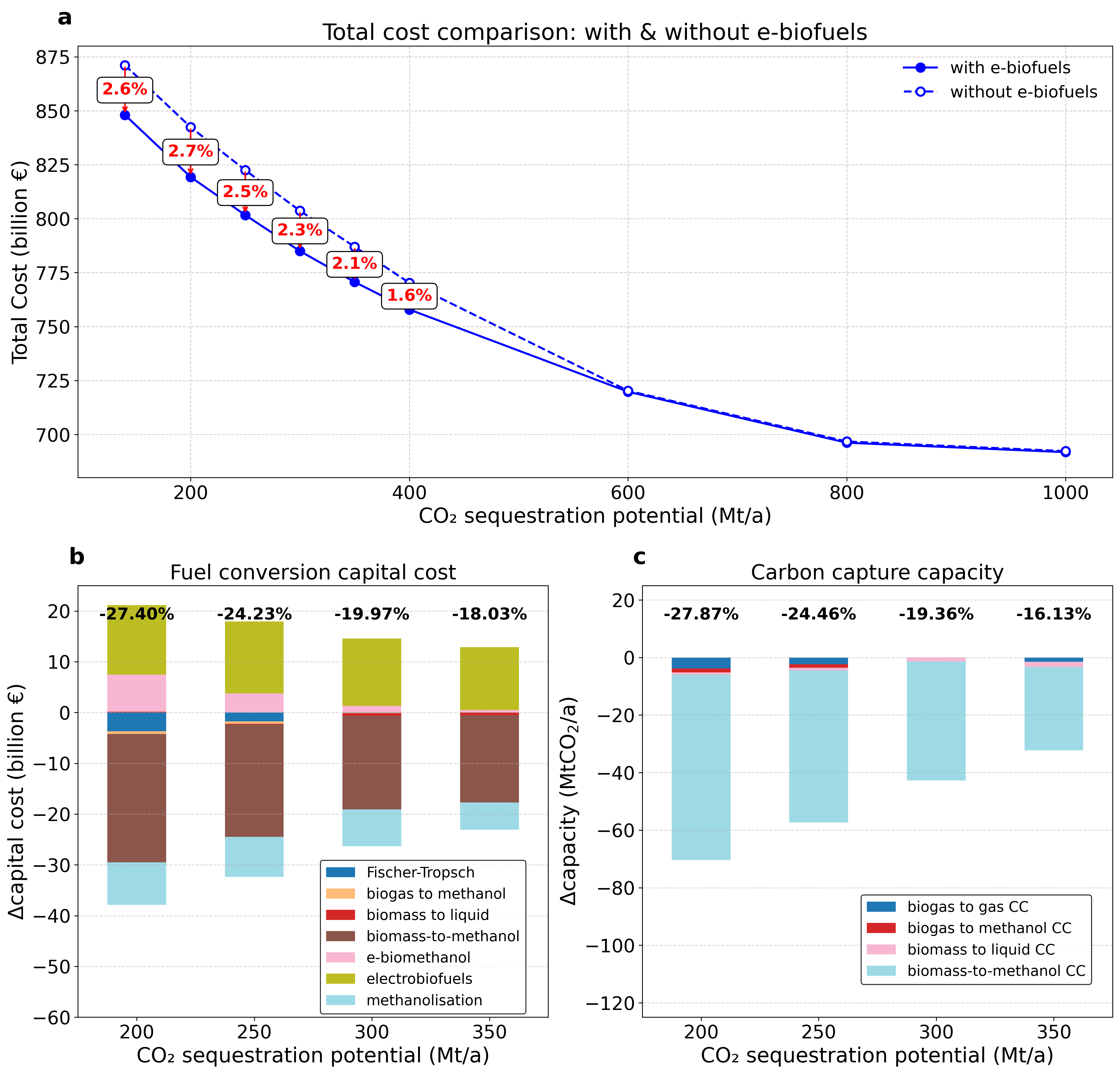}
    \caption{Effects of including e-biofuels on total system costs, fuel-conversion capital, and carbon-capture capacity under different levels of carbon-sequestration potential. Here, $\Delta$ denotes the difference between the scenario \textit{with e-biofuels} and the scenario \textit{without e-biofuels} (i.e., capacity$_{\text{with}}$ $-$ capacity$_{\text{without}}$, capital$_{\text{with}}$ $-$ capital$_{\text{without}}$).}
    \label{fig:costDiff_nopolicy}
\end{figure}

E-biofuels become an increasingly competitive liquid fuel supply option the more fossil fuels are phased out. Fossil fuel usage in a net-zero energy system is affected by renewables targets in hard-to-abate sectors, and the extent to which carbon sequestration (CS) is made available by scaling up geological storage. Biofuels (without hydrogen addition) are a cost-effective part of the fuel mix over a wide range of CS availability, covering up to 46\% of liquids demand (Fig.~\ref{fig:fuelmix_CSsweep}). At higher CS availability, capturing and storing excess carbon from biofuel production and other processes is cost-effective, but the scarcer CS capacity is, the more valuable utilisation of excess carbon becomes (Fig.~\ref{fig:S3_el_bio_CC_sweep}b). It is more cost-effective to directly utilise excess carbon through hydrogen addition, rather than capturing, transporting, and separately synthesising fuels, and at CS capacities of 300 Mt and below, e-biofuels are the dominant liquids supply option  (Fig.~\ref{fig:fuelmix_CSsweep}a). Integrating e-biofuels reduces required fuel conversion capital by up to 27\% (Fig.~\ref{fig:costDiff_nopolicy}b), carbon capture capacity by 28\% (Fig.~\ref{fig:costDiff_nopolicy}c), and total system cost by 23~B\texteuro/a (2.7\%) (Fig.~\ref{fig:costDiff_nopolicy}a) at high renewable shares, compared to the system without e-biofuels. Most of the cost reduction originates from the liquid fuel system, with a decrease of 42~B\texteuro/a (11\%) (Fig.~\ref{fig:S3_liquid_total_cost_nopolicy}), partly offsetting cost increases in other parts of the energy system. Inclusion of e-biofuels also reduces the \co \ price by up to 16\% (Fig.~\ref{fig:S2_CO2_Price}). It makes it more cost-effective to utilise excess carbon, thereby dampening biomass demand by up to 221 TWh, equivalent to nearly 19\% of Europe’s assumed domestic residue potential(Fig.~\ref{fig:S3_el_bio_CC_sweep}a).

\clearpage
\section*{Carbon system value governs the liquid-fuel merit order}

\begin{figure}[!htp]
    \centering
    \includegraphics[width=1\linewidth]{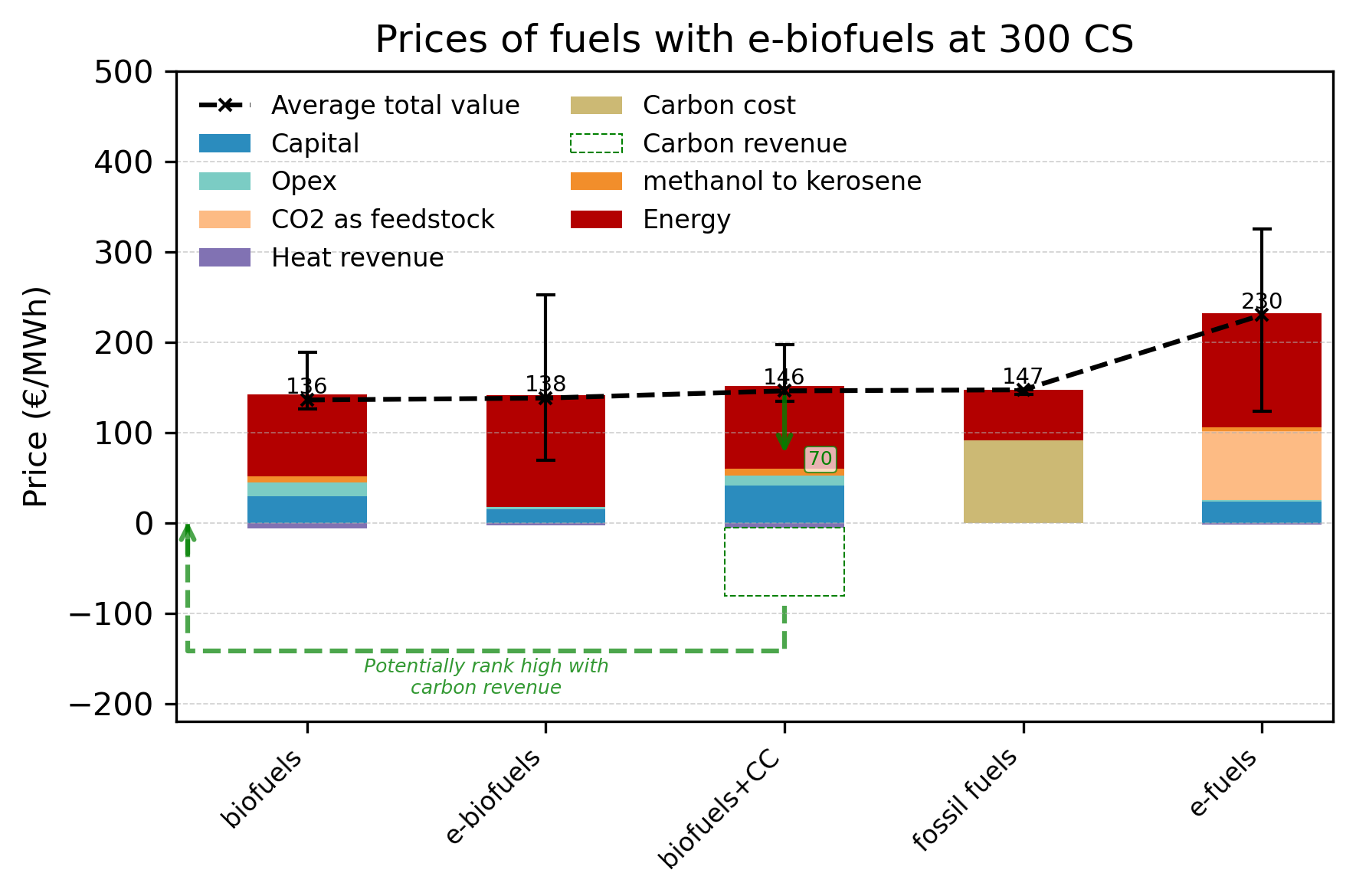}
    \caption{
        Merit order of liquid fuels using in aviation under baseline scenarios with carbon sequestration (CS) potentials of 300~Mt~a$^{-1}$. The upper and lower bounds indicate the range of system costs of different technologies across different time steps and spatial nodes, reflecting temporal and spatial heterogeneity in technology costs. For biofuels with carbon capture pathways (biofuels+CC), carbon-removal revenues could be included when the captured \co\ is permanently stored underground. This category is therefore shown with dashed bars, as its effective revenue depends on the availability and utilisation of carbon sequestration capacity, which can potentially raise its position in the merit order.
    }
    \label{fig:merit_order_nopolicy}
\end{figure}

By valuing each liquid-fuel pathway using system shadow prices which reflect scarcity rent, we show that the system value of carbon governs the liquid fuel merit order and explains the deployment patterns observed above (Fig.~\ref{fig:merit_order_nopolicy}). Carbon sequestration availability strongly affects the feasibility of using fossil fuels and the shadow price of carbon. With limited carbon sequestration, conventional biofuels are cost-competitive on average (Fig.~\ref{fig:merit_order_nopolicy}), but spatial and temporal heterogeneity in hydrogen value renders a large fuel cost span to make e-biofuels frequently more cost-effective, whereas e-fuels are consistently disadvantaged by the additional cost of sourcing \co\ feedstock despite similar hydrogen requirements.

A higher carbon sequestration capacity enables more fossil fuel usage offset by negative emissions (Fig.~\ref{fig:S4_merit_order_300_600CS}) and decreases \co \ prices (Fig.~\ref{fig:S2_CO2_Price}e), whereby the competitiveness of using hydrogen to produce e-fuels and e-biofuels decreases.

\clearpage
\section*{E-biofuels in European renewable fuel mandates}

\begin{figure}[!htp]
    \centering
    \includegraphics[width=1\linewidth]{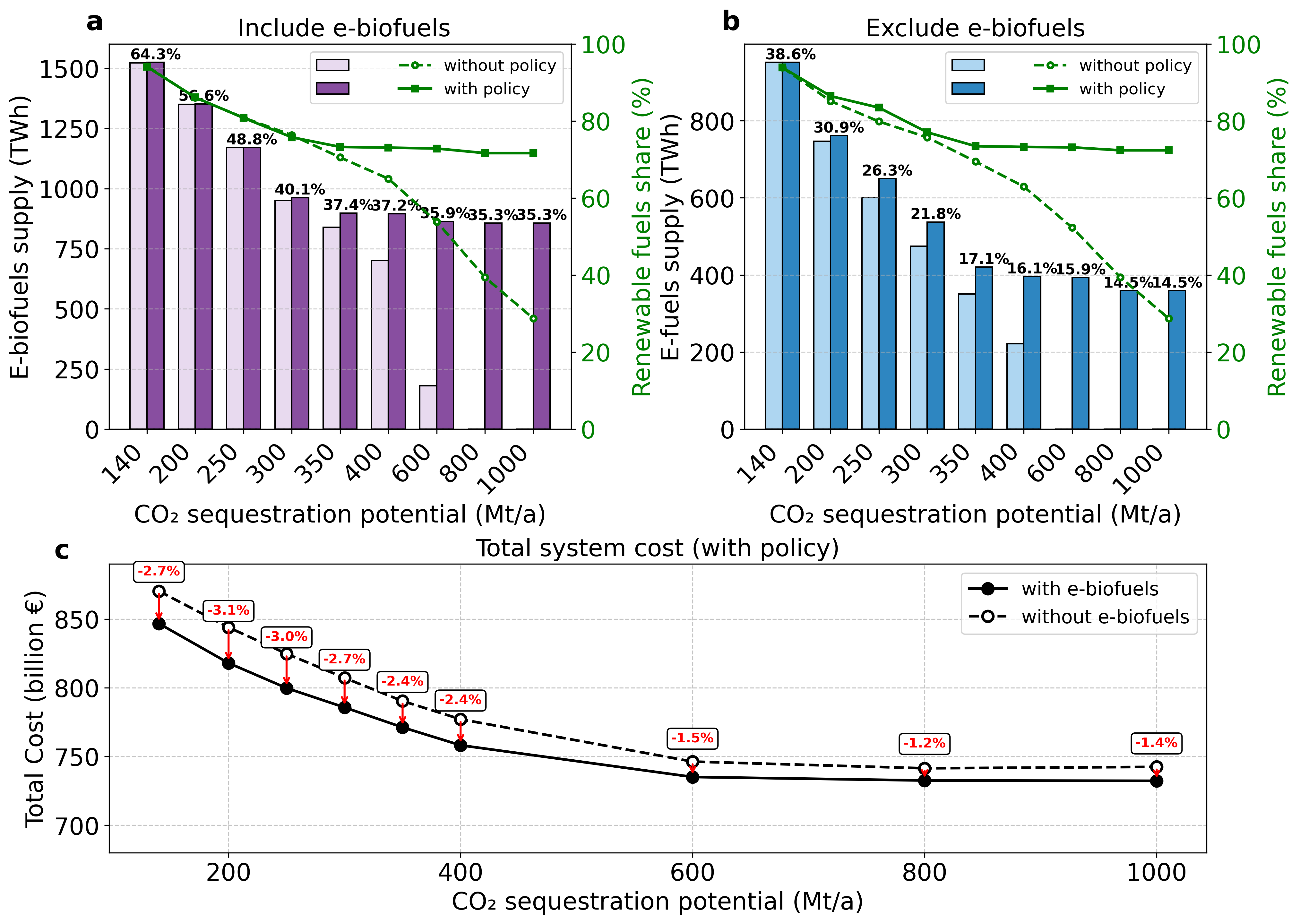}
    \caption{E-biofuels fuel supply and system cost comparison for different scenarios. Panel (a) shows the impact of introducing policy constraints on e-biofuel deployment in scenarios with e-biofuels. Panel (b) shows the impact of introducing policy constraints on e-fuel deployment in scenarios without e-biofuels. The percentages above the bars in Panels (a) and (b) denote the share of e-biofuel and e-fuel in the total liquid fuel mix under policy constraint. Panel (c) compares total system costs between the scenarios with and without e-biofuels when there are renewable-fuel policy constraints.}
    \label{fig:fuels_mix_policy}
\end{figure}

Introducing European renewable fuel mandates has a negligible effect on system cost and fuel composition at carbon sequestration capacities of 250 Mt \co/a and below (Fig.~\ref{fig:fuels_mix_policy}a), as e-biofuels and biofuels already supply more than 80\% of aviation liquid-fuel demand in these cases (Fig.~\ref{fig:S5_two_sector_cost}). At higher CS availability, the mandates become increasingly binding, raising the renewable fuel share from 23\% at the higher end of carbon sequestration capacity to over 72\%. Thereby, the mandate reduces investor uncertainty to how much fossil fuels are permissible.

Without e-biofuels, compliance is achieved mainly through e-fuels, whose deployment increases to about 15\% (Fig.~\ref{fig:fuels_mix_policy}b). When e-biofuels are available, by contrast, they become the dominant compliance option even at high CS availability, maintaining a supply share above 35\%, displacing most e-fuels and part of conventional biofuels, while keeping system costs 1.4–3.1\% lower than in scenarios without e-biofuels (Fig.~\ref{fig:fuels_mix_policy}a–c).

\clearpage
\section*{Sensitivity of e-biofuels to system and policy uncertainties}

\begin{figure}[!htp]
    \centering
    \includegraphics[width=1\linewidth]{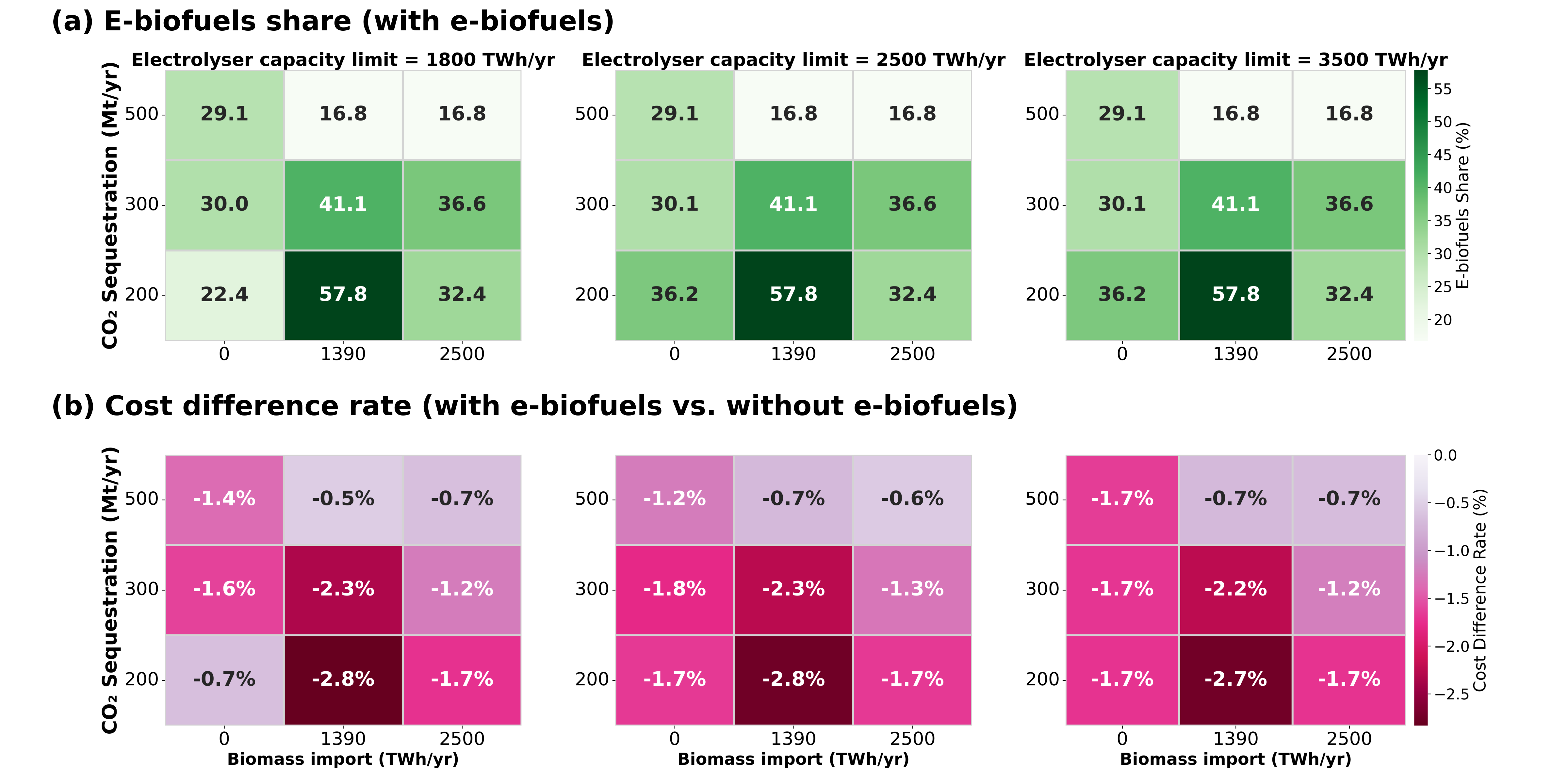}
    \caption{\textbf{Sensitivity of e-biofuels deployment and system cost to biomass import, electrolyser capacity and CO$_2$ sequestration potential.} Panel (a) shows the e-biofuels share (\%) under the \textit{with e-biofuels} setting, and panel (b) shows the system cost difference rate (\%) between \textit{with e-biofuels} and \textit{without e-biofuels}. Each panel reports results for three electrolyser capacity limits (1800, 2500, and 3500~TWh/a). 
    The x-axis is biomass import (0, 1390, 2500~TWh/a), and the y-axis is CO$_2$ sequestration potential (200, 300, 500~Mt/a). Negative values in panel (b) indicate cost reductions when e-biofuels are available. Figures show the total system costs decline with increasing CO$_2$ sequestration capacity. The marginal cost reduction from additional biomass imports decreases beyond moderate import levels. Increasing electrolyser capacity reduces system costs primarily under biomass- and CO$_2$-constrained conditions, whereas additional hydrogen capacity has limited impact when either biomass availability or CO$_2$ sequestration capacity is abundant.}
    \label{fig:biomass_CS_sensitivity}
\end{figure}

Sensitivity analyses first explore the effects of varying biomass import availability, H$_2$ resources, and \co \ sequestration potential (Fig.~\ref{fig:biomass_CS_sensitivity}). E-biofuels attain sizeable deployment shares (up to 57.8\%) and are consistently associated with lower total system costs, yielding reductions of up to 2.8\% relative to the case without e-biofuels. The cost benefit is strongest when biomass availability is moderate and \co \ storage is limited (biomass imports comparable to domestic availability and \co \ sequestration not exceeding the 2040 EU target), whereas additional biomass imports beyond this level provide diminishing returns with more biofuels deployment. 

\begin{figure}[!htp]
    \centering
    \includegraphics[width=1\linewidth]{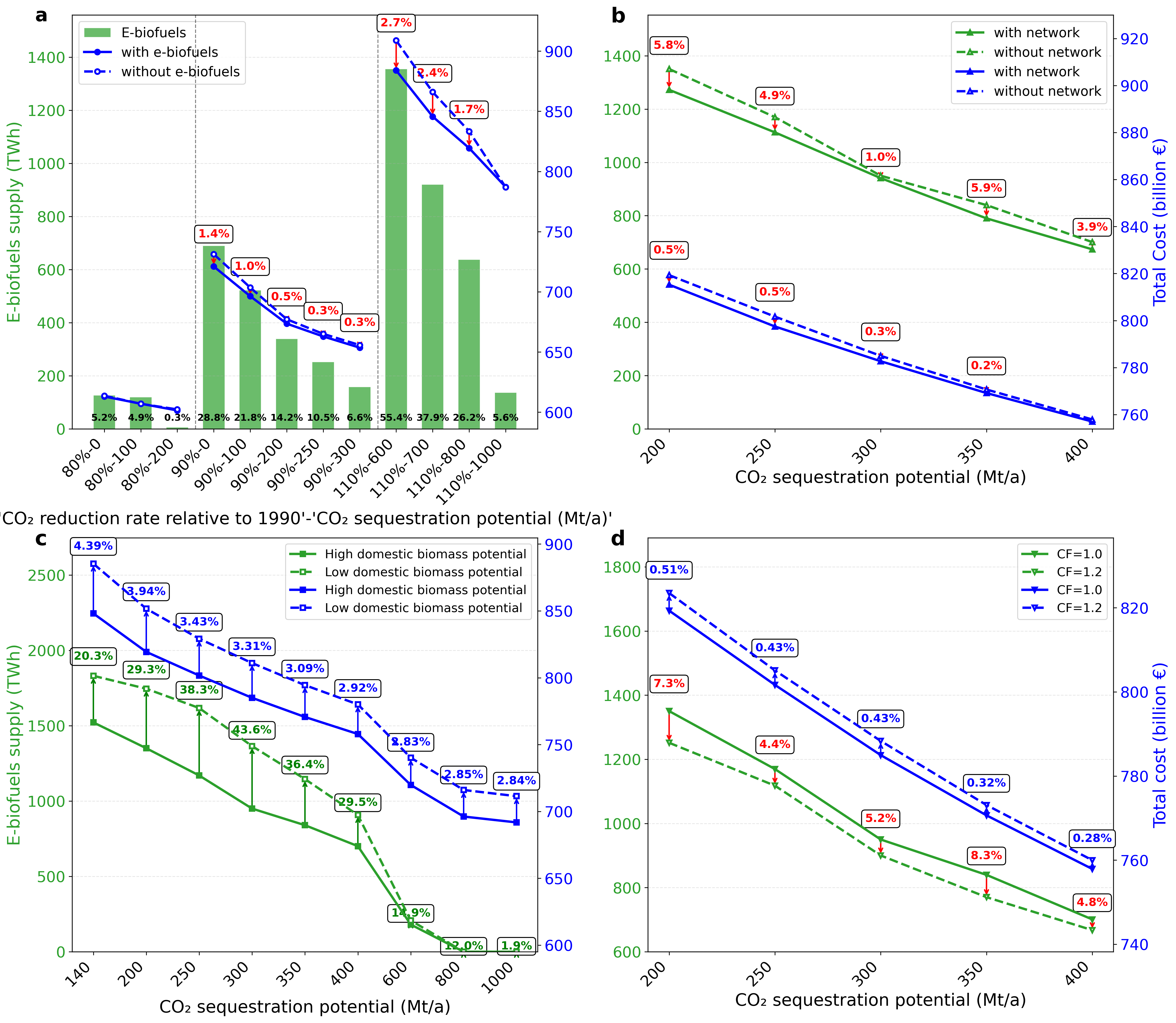}
    \caption{Sensitivity analysis under different carbon sequestration potentials. Panel (a) shows e-biofuel deployment and total system cost under different emissions-reduction targets and carbon-sequestration (CS) availability, comparing scenarios with and without e-biofuels. The black numerical labels above the horizontal axis represent the proportion of the deployment volume of e-biofuels to the total deployment volume of liquid fuels. Panel (b) compares e-biofuel deployment and total system cost in the presence and absence of \co \ and hydrogen transport networks. Panel (c) illustrates e-biofuel deployment and total system cost under medium versus low domestic biomass availability. Panel (d) shows the impact of a 20\% increase in e-biofuel capital investment costs on e-biofuel deployment and total system cost.}
    \label{fig:carbon_target_sensitivity}
\end{figure}

Renewable fuels become cost-effective only at high emissions targets to mitigate the last 10\%, with a negligible role at a -80\% target (Fig.~\ref{fig:carbon_target_sensitivity}a). However, e-biofuels reach substantial deployment levels already at a –90\% target, reducing system cost by up to 1.4\%, and account for cost-effective supply shares of up to 55\% of liquid fuel demand under \textminus110\% net-negative targets, reducing system cost by up to 2.7\%. 

If biomass is restricted to the medium domestic residue potentials in JRC ENSPRESO \cite{ruiz2019enspreso}, e-biofuels reduce system costs by up to a substantial 4.4\% at a net-zero target compared to the low residue potential, with cost-effective supply shares increasing by up to 44\% (Fig.~\ref{fig:carbon_target_sensitivity}c).

If, however, \co \ and hydrogen infrastructure are co-optimised, the effect of including e-biofuels decreases; their deployment share decreases by up to 6\%, but the system cost only reduces by at most 0.5\% compared to baseline scenarios lacking both \co \ and hydrogen networks (Fig.~\ref{fig:carbon_target_sensitivity}b).

The role of e-biofuels is insensitive to investment-cost uncertainty. A 20\% increase in capital costs for e-biofuel production technologies (ceteris paribus) reduces deployment by only about 4\% and raises total system costs by 0.5\% (Fig.~\ref{fig:carbon_target_sensitivity}d), indicating limited sensitivity to plausible cost variations even when neglecting the correlation of the cost of other fuel pathways with e-biofuel costs.

\clearpage
\section*{Discussion}
Liquid fuels and chemicals are among the hardest to replace in the transition towards a net-zero energy system \cite{Millinger2025,schreyer2025net}. As shown in the sensitivity analysis, renewable liquids become a significant part of the cost-effective mix only at emissions targets of -90\% and above, however the EU is proposing to achieve across all sectors already by 2040 \cite{eu2025climate_law_amendment}. More stringent carbon emission targets in the EU energy sector may be needed to compensate for residual emissions from agriculture and the land-use, land-use change, and forestry (LULUCF) sector. The availability of sufficient renewable fuel capacities by then requires the establishment of complex and novel value chains. Many uncertainties, ranging from electrolyser deployment and green hydrogen availability to carbon capture, carbon sequestration and biomass supply, continue to inhibit investment. At the same time, infrastructure and markets need time to develop and mature in order to deliver at scale when required. This raises the question of whether more flexible and adaptive fuel production pathways can help mitigate these risks. 

Although the REPowerEU strategy sets an ambition to produce renewable hydrogen equivalent to around 10\% of the EU’s final energy demand by 2050 (approximately 1400 TWh) \cite{EC_REPowerEU_2022}, a level comparable to the hydrogen supply in our baseline scenario under 300 Mt\co/a \ sequestration (Fig.~\ref{fig:S3_el_bio_CC_sweep}a), the scale-up of electrolysers is subject to substantial inertia \cite{Odenweller2022}. Moreover, deploying green hydrogen to replace existing fossil-based hydrogen, or using it directly in applications such as the direct reduction of iron, achieves greater cost-effectiveness and emissions abatement than converting hydrogen into e-fuel synthesis \cite{Ueckerdt2021}. This implies that hydrogen-based liquid fuels are likely to remain constrained in the near to medium term. Substantial expansion of both renewable power generation and electrolysis capacity is therefore required before hydrogen can play a major role in e-fuel or e-biofuel production. 

At the same time, it is also uncertain how fast CCS can scale up. The EU plans to increase \co \ injection capacity from current 2 Mt\co/a to at least 50 Mt\co/a by 2030 and 250 Mt\co/a from 2040 onwards under the Net-Zero Industry Act \cite{eu2024reg1735}. Given high failure rates of historical CCS endeavours \cite{Kazlou2024} and potentially strong limitations for geological sequestration \cite{Gidden2025} and \co \ injection rates \cite{Grant2022}, strategies assuming high CCS deployment carry substantial risk, as do strategies relying on high renewable fuel shares. These uncertainties need to be taken into account when planning the transition in sectors relying on liquid carbonaceous molecules.

In contrast, substantial biomass residue resources remain untapped and could be utilised to increase biofuel production for emissions abatement in the near term. As the energy transition progresses, e-biofuels, conceived as a modular production system, enable an incremental investment approach that can then build upon installed biofuel infrastructure, when there is more certainty about the availability of green H$_2$ and carbon sequestration, and whether it is more economical to utilise or store excess carbon. If green H$_2$ remains scarce but carbon sequestration scales up, excess carbon can be transported and permanently stored. For syngas technologies, acid gases such as excess \co \ need to be separated anyway and the additional effort would be to transport it to where it can be stored. If on the other hand carbon sequestration remains limited but green H$_2$ scales up, biofuel processes can be modified to enable injection of H$_2$ to enhance carbon efficiency and produce e-biofuels, without the need for substantial additional investments. Such incremental investment strategies creates significant real option value by providing the flexibility to 'wait and see' without delaying climate action.

While existing studies show that the transition to low-carbon liquid fuels involves a fundamental resources trade-off between biomass, renewable electricity, hydrogen, and geological carbon sequestration \cite{Mignone2024,schreyer2025net}, how the e-biofuels would affect this nexus has not yet been evaluated. Our results show that introducing e-biofuels qualitatively reshapes the regime structure of dominant liquid-fuel supply across the biomass-import and \co \ sequestration space (Fig.~\ref{fig:biomass_CS_h2_sensitivity}): biofuels, e-fuels and fossil fuels dominate mainly in the different 'corner' cases (high biomass for biofuels; low biomass and low CS for e-fuels; low biomass and high CS for fossil fuels), whereas e-biofuels dominate under more intermediate conditions (biomass imports comparable to domestic availability and \co \ sequestration not exceeding the 2040 EU target). This pattern indicates that e-biofuels can act as a hedge against simultaneous resource constraints by drawing on a more diversified resource base, thereby reducing reliance on any single critical input and strengthening robustness to supply disruptions and spatio–temporal mismatches across biomass, electricity (or H$_2$) and carbon storage. 

Beyond this resource-hedging role, e-biofuel systems also carries broader strategic implications for managing policy uncertainty throughout the energy transition. By enabling a sequential evolution of the fuel production mix between the present and an uncertain future, they support a gradual investment response while reducing the risk of stranded assets. This flexibility provides a hedge against policy uncertainty, as the EU \textminus90\% target is subject to renegotiation, making it less reliable especially for hard to replace options that represent the last 10\% towards net-zero. In the longer term, such a sequential evolution also enables a shift away from biomass use, should DAC become a competitive option. This potential shift increases robustness against delays for or absence of bulky infrastructure, such as \co \ and H$_2$ transport and storage networks. In this sense, e-biofuel systems can serve as a transitional platform that remains compatible with both biomass-based and DAC-based carbon supply chains.

Realising this potential depends on the maturity and scalability of key technologies. From a technological perspective, the principal scale-up risks of e-biofuels, as with syngas-based biofuels, stem from biomass gasification and syngas conditioning, which remain less mature and harder to standardise than downstream hydrogenation, Reverse Water Gas Shift (RWGS) and fuel synthesis units. Recent reviews highlight that while modern gasifiers show promising conversion efficiencies and potential for integration with green hydrogen, substantial improvements in reactor design, tar mitigation and feedstock flexibility are still needed for industrial readiness \cite{kumar2025advancements}. Syngas cleaning and conditioning further complicate scale-up due to the complexity of biomass composition and the engineering integration required to meet the specifications of Fischer–Tropsch or methanol synthesis \cite{patel2025review}. However, hydrogen production and downstream synthesis technologies are modular and commercially mature, with current maximum capacities of around 0.2 Mt/a and 0.6 Mt/a, respectively, enabling incremental expansion as renewable electricity becomes available. Compared with syngas-based biofuels, e-biofuels reduce scale-up pressure on gasification and syngas cleaning by enhancing and simplifying syngas composition through hydrogen addition.

Of globally planned SAF projects, only around 10\% are gasification-based, yet roughly 60\% of these already plan to integrate hydrogen addition, i.e. e-biofuels. 72\% of SAF capacity relies on HEFA (Hydroprocessed Esters and Fatty Acids) pathways with limited feedstock potential \cite{genasolutions2025}. In the EU (including UK and Norway), announced SAF capacity reaches about 10 Mt/a by 2030, comprising roughly 10\% gasification-based SAF (of which 80\% are from co-processing) and 20\% e-SAF \cite{martulli2025potential}. Our results indicate that, under a carbon-sequestration capacity of 250 Mt~\co/a, a cost-effective configuration for meeting European renewable-fuel mandates in aviation includes the expansion of e-biofuels and biofuels corresponds to an average compound annual growth rate (CAGR) of around 24\%/a and 26\%/a between 2030 and 2050, respectively. These values are comparable to the interquartile range of growth rates observed during the fastest historical expansion phases of biofuels and ethanol in Europe (17\%-27\%/a) \cite{martulli2025potential}. By contrast, excluding e-biofuels leads the system to rely on renewable fuel mixes dominated by conventional biofuels and e-fuels, with average CAGR of up to 30\% and 15\% by 2050, respectively. The former approaches the acceleration rates historically observed during rapid scale-up of wind and solar power in Europe (31\%-54\%/a) \cite{martulli2025potential}. Which pathway is most likely to emerge is difficult to assess at this stage. On the one hand, the actual realization rate of announced capacities by 2030 remains highly uncertain; on the other hand, while the integration of e-biofuels demands a lower absolute growth rate for meeting mandates, the inherent modularity of technologies like DAC and e-fuels may facilitate faster scaling, as observed  in solar, compared to the more customized, site-specific nature of (e-)biofuel infrastructure.

So-called second best policies and policy mixes can provide more market stability and be near-optimal next to the ideal \co \ pricing \cite{Dimanchev2025}. Scaling up renewable fuel markets through mandates such as implemented with ReFuelEU Aviation \cite{EU2023_aviation} provides more investment certainty, if market actors can trust the policy to remain, and if it is clear which fuels are eligible. The Renewable Energy Directive (RED) lacks clear clarifications and concrete examples regarding Renewable Fuels of Non-Biological Origin (RFNBOs)\cite{Pettersson_RISE_REDFuels_2025}, such as e-biofuels, which are not explicitly mentioned in RED. A more in-depth analysis of RED, however, indicates that co-processing can be applied to e-biofuels, which therefore constitute a blend of biofuels and RFNBOs (the fraction derived from hydrogen) \cite{EU_DelegatedReg_2023_1185,EC_HydrogenDelegatedActs_QA_2024}. In ReFuelEU Aviation, the impression is given that synthetic fuels refer exclusively to e-fuels, i.e., fuels produced from hydrogen and captured carbon dioxide, and thus do not include e-biofuels: “Synthetic aviation fuels from renewable hydrogen and captured carbon (in the meaning of Article 2(36) of RED and limited to liquid drop-in fuels only).” However, reference is made to Article 2(36) of RED, which defines RFNBOs, a category under which the hydrogen-derived component of e-biofuels falls. Still, clear formulations in legislation regarding the potential role of e-biofuels in meeting renewable fuel and RFNBO mandates are desirable, as they can reduce investment uncertainty and enable a more cost-effective pathway toward achieving policy targets. Furthermore, current legalisation favours the less energy-efficient option of e-fuels (classified as 100\% RFNBO) over e-biofuels (where about 2/3 could be RFNBO depending on how much hydrogen is used), thereby creating a lock-in in a sub-optimal system \cite{Pettersson_RISE_REDFuels_2025,beiron2026locked}. A change in current legislation has therefore been suggested, to create a more consistent and technology-neutral classification of fuels \cite{Pettersson_RISE_REDFuels_2025}. 

Taken together, our findings suggest that e-biofuels can serve as a cost-effective and strategically resilient bridge in the transition to net-zero liquid fuels. Unlocking this potential, however, depends on well-designed policy frameworks, including clearer regulatory treatment of e-biofuels, improved transparency and consistency in fuel classification, and policy incentives that better reflect overall system efficiency. Without such adjustments, regulatory ambiguity and misaligned incentives risk delaying the deployment of otherwise viable pathways, ultimately increasing the cost and uncertainty of reducing the climate impact of the most difficult segments of the energy system.

\section*{Methods}

\subsection*{PyPSA-Eur model}

PyPSA-Eur \cite{brown2018synergies,brown2026_pypsaeur} is an open-source, sector-coupled, full European energy system optimization model based on linear programming. It encompasses the power sector, transport (including shipping and aviation), space and water heating, industry, and industrial feedstocks. The model minimizes total system costs by co-optimizing capacity expansion and operation of all energy generation, conversion, storage, and transmission technologies across electricity, hydrogen, and gas. It is built on the Python software toolbox PyPSA (Python for Power System Analysis) \cite{brown2026_pypsaeur}. A comprehensive model description is provided in Neumann \textit{et al.} \cite{neumann2023potential}.

A 39-node spatial resolution and a 3-hourly temporal resolution were applied over a full year in overnight greenfield scenarios, chosen as a trade-off between computational effort and the difference in results compared to a 1-hour resolution \cite{Millinger2025}. This European energy system includes EU-27 member states, UK, Norway, Switzerland and the Balkans. A country-level spatial resolution was adopted for computational feasibility. A lossy electricity transmission model was employed, which is appropriate at this level of aggregation \cite{neumann2022assessments}, and transmission expansion was constrained to at most 1.5 times the total line volume in 2022.

Final energy demands across sectors are derived from the JRC IDEES database \cite{mantzos2017jrc}, with additional data for non-EU countries (see refs.~\cite{victoria2022speed,zeyen2021mitigating} for further details). These demands are perfectly inelastic and must be met in all model scenarios. In contrast, the production of energy carriers, including electricity, hydrogen, methane, and liquid fuels, is determined endogenously. Fossil fuels (coal, natural gas, and oil) and uranium are included, while coal with carbon capture is excluded. Medium domestic (country-level) biomass residue potentials for 2040 were obtained from the JRC ENSPRESO database \cite{ruiz2019enspreso}, explicitly represented at the nodal level. A weighted average of country-specific biomass costs was used, including harvesting, collection, and transport, based on the reference biomass scenario \cite{jrc_enspreso2021}. Additionally, imports of more expensive solid biomass are allowed \cite{millinger2022biofuel}, while potential upstream emissions are not considered.

Technology costs and efficiencies are described in the dataset repository \cite{zeyen2026_technologydata_ebiofuel}. Main values for 2040 (in \texteuro$_{2015}$) are taken from the PyPSA energy system technology dataset v0.12.0. A uniform discount rate of 7\% is applied across all countries. The version of the model used in this study (v2025.04.0) integrates recent developments, including additional competition for industrial heat supply \cite{Millinger2025}, the utilization of waste heat from different synthetic fuels production process, and expanded methanol production routes (e.g., grey methanol, blue methanol, and biogas-to-methanol), and methanol utilization pathways (e.g., methanol-to-HVC) \cite{glaum2025minMeOH}. Most importantly, four types of e-biofuels were newly introduced: e-biomass-to-liquid, e-biomethanol, e-biogas-to-methanol, and e-bioSNG.

\subsection*{E-biofuels in PyPSA-Eur}

To represent e-biofuel production within PyPSA-Eur, we formulate a thermodynamically consistent carbon–energy allocation framework. The Fischer–Tropsch (FT) pathway is used as a representative synthesis route to illustrate the underlying principles. The same carbon accounting logic is subsequently generalised to other hydrogen-assisted biomass upgrading pathways.

\subsection*{Carbon balances of biomass-to-fuel conversion}

We define the specific CO$_2$ emission intensities of biomass and liquid fuel [tCO$_2$/MWh] as $\varepsilon_{bio}$ and $\varepsilon_{fu}$, respectively.

Using the molar mass ratio between carbon and CO$_2$ ($m_C/m_{CO_2} = 12/44$), the corresponding carbon intensities [tC/MWh] are

\begin{equation}
c_{bio} = \frac{m_C}{m_{CO_2}} \varepsilon_{bio}
\end{equation}

\begin{equation}
c_{fu} = \frac{m_C}{m_{CO_2}} \varepsilon_{fu}
\end{equation}

Let $E_{bio}$ (MWh) denote biomass energy input and $E_{fu}$ (MWh) denote liquid fuel energy output.

The biomass-to-fuel energy efficiency is defined as

\begin{equation}
\eta_{bio} = \frac{E_{fu}}{E_{bio}}
\end{equation}

The carbon efficiency, defined as the fraction of biomass carbon retained in the final fuel, is

\begin{equation}
\eta_c =
\frac{c_{fu} E_{fu}}
     {c_{bio} E_{bio}}
\end{equation}

Substituting the definitions above yields

\begin{equation}
\eta_c
=
\eta_{bio}
\frac{\varepsilon_{fu}}
     {\varepsilon_{bio}}
\end{equation}

which can be rearranged as

\begin{equation}
\eta_{bio}
=
\eta_c
\frac{\varepsilon_{bio}}
     {\varepsilon_{fu}}
\end{equation}

\paragraph{System with hydrogen addition}

When electrolytic hydrogen is supplied, part of the carbon that would otherwise be emitted as CO$_2$ is converted into additional fuel. Biomass input $E_{bio}$ remains constant.

Let $E_{fu}'$ (MWh) denote total fuel production after hydrogen addition. The updated biomass-to-fuel efficiency becomes

\begin{equation}
\eta_{bio}' = \frac{E_{fu}'}{E_{bio}}
\end{equation}

and the updated carbon efficiency is

\begin{equation}
\eta_c' =
\frac{c_{fu} E_{fu}'}
     {c_{bio} E_{bio}}
\end{equation}

The increase in carbon efficiency is

\begin{equation}
\Delta \eta_c = \eta_c' - \eta_c
\end{equation}

From carbon conservation, the additional fuel production enabled by hydrogen is

\begin{equation}
\Delta E_{fu}
=
\frac{\Delta \eta_c}{\eta_c'} E_{fu}'
\end{equation}

\paragraph{Hydrogen-related efficiencies}

Let $E_{H_2}$ (MWh) denote hydrogen energy input.

The Fischer–Tropsch efficiency is defined as incremental fuel production 
per unit hydrogen input:

\begin{equation}
\eta_{FT}
=
\frac{\Delta E_{fu}}{E_{H_2}}
\end{equation}

The hydrogen-to-fuel efficiency is defined as total fuel production 
per unit hydrogen input:

\begin{equation}
\eta_{H_2}
=
\frac{E_{fu}'}{E_{H_2}}
\end{equation}

Substituting the expression for $\Delta E_{fu}$ yields

\begin{equation}
\eta_{FT}
=
\frac{\Delta \eta_c}{\eta_c'}
\eta_{H_2}
\end{equation}

or equivalently,

\begin{equation}
\eta_{H_2}
=
\frac{\eta_c'}{\Delta \eta_c}
\eta_{FT}
\end{equation}

\subsection*{Extension to other hydrogen-assisted biomass/biogas pathways}

Although the above derivation is formulated for FT-based e-biofuel production, the carbon–hydrogen co-conversion principle is general and can be extended to other biomass or biogas upgrading routes.

In addition to FT diesel, this study considers three further pathways:

\begin{itemize}[noitemsep,leftmargin=*]
    \item \textbf{e-bioSNG}: biomass-derived syngas upgraded via methanation;
    \item \textbf{e-biomethanol}: biomass-derived syngas upgraded via methanol synthesis;
    \item \textbf{e-biogas-methanol}: upgraded biogas followed by methanol synthesis.
\end{itemize}

For a generic e-biofuel pathway $x$, total fuel production is expressed as

\begin{equation}
E_{fu}^{(x)}
=
\eta_{bio}^{\prime (x)}
E_{bio}^{(x)}
=
\eta_c^{\prime (x)}
\frac{\varepsilon_{bio}^{(x)}}
     {\varepsilon_{fu}^{(x)}}
E_{bio}^{(x)}
\end{equation}

The overall hydrogen–biomass co-conversion efficiency (relative to total primary energy input) is formulated as

\begin{equation}
\eta_{total}^{(x)}
=
\frac{E_{fu}^{(x)}}
     {E_{bio}^{(x)} + E_{H_2}^{(x)}}
\end{equation}

Using the definitions of $\eta_{bio}'^{(x)}$ and $\eta_{H_2}^{(x)}$, 
this can be rewritten as

\begin{equation}
\eta_{total}^{(x)}
=
\frac{1}
{\frac{1}{\eta_{bio}^{\prime (x)}} 
 + \frac{1}{\eta_{H_2}^{(x)}}}
\end{equation}

\subsection*{Parametrisation of e-biofuel technologies}
To represent the cost and performance characteristics of hydrogen-assisted biofuel technologies, we parameterise e-biomass-to-liquid, e-biomethanol, e-bioSNG, and e-biogas-methanol based on published techno-economic studies. The methodological assumptions for each pathway are summarised as follows.

For e-biomass-to-liquid, the specific capital investment and operating expenditures are assumed follow research \cite{hillestad2018improving}. Hydrogen addition substantially enhances carbon utilisation: the overall carbon efficiency increases from 0.38 to 0.91. This system comparison assumed a plant capacity of 435 MW$_{th}$ based on the LHV of dry biomass.

For e-biomethanol, the capital investment and variable operating costs are sourced from literature \cite{anetjarvi2023benefits}. Hydrogenation increases the overall carbon efficiency from 0.45 to 0.9. The cost estimation results for cases of 100 ktons/a of methanol production.

For e-bioSNG, the specific cost data is from  \cite{hannula2016hydrogen}. Hydrogen addition raises the overall carbon efficiency from 0.58 to 0.98. The data is estimated based on a plants of 100MW (LHV) biomass input. 

For e-biogas-methanol, capital and operating costs are set based on simulation results of \cite{park2025bio}. Carbon efficiency increases markedly from 0.616 to 0.885. The system was assessed at a commercial scale of 2000 t/d, the typical scale of modern methanol plants.

Across all e-biofuels configurations, hydrogen is used to upgrade the carbon utilisation efficiency through 
enhanced syngas rebalancing and improved Fischer-Tropsch or methanol synthesis stoichiometry. Cost differences on different system process towards increased total production, are mainly from (i) additional cost for reversed water gas shift system and capacity expansion, and (ii) increased plant complexity, and (iii) reduced water gas shift processes or air separate units to supply O$_2$. Energy and carbon efficiencies represent the ratio of final fuel energy or utilised biogenic carbon relative to the biomass, electricity and hydrogen inputs. 
Overall, the parameter settings reflect a consistent modelling framework in which hydrogen-assisted upgrading routes enable substantial improvements in carbon utilisation and fuel yields, at the cost of higher electricity and hydrogen demand compared to routes without H$_2$ addition. These assumptions ensure comparability of techno-economic performance across the different e-biofuel pathways.

\subsection*{EU policy constraints}
The ReFuelEU Aviation regulation introduces a mandatory Sustainable Aviation Fuel (SAF) blending requirement for aviation fuel in the EU, starting at 2\% in 2025 and gradually increasing to 70\% by 2050. Within this target, a dedicated sub-mandate is set for Renewable Fuels of Non-Biological Origin (RFNBOs), rising from 1.2\% in 2030 to 35\% in 2050. Eligible fuels include SAF that meet RED sustainability criteria, notably advanced biofuels produced from waste and residue inputs, biofuels derived from oils and fats, recycled carbon fuels, as well as RFNBOs. In addition, the framework allows the use of synthetic low-carbon aviation fuels and low-carbon hydrogen as compliance options, provided they achieve at least 70\% lifecycle emissions reduction, enabling flexibility while gradually steering the sector toward fully renewable, electricity-based e-fuels in the long term.

The ReFuelEU Maritime policy is the central framework for shipping within Fit for 55, targeting an 80\% reduction in the greenhouse gas intensity of marine fuels by 2050 compared to 2020, with reductions introduced progressively starting in 2025. Eligible alternative options under this policy include green methanol, ammonia, hydrogen-derived RFNBOs, advanced biofuels, and future synthetic or bio-LNG, with shore-side electrification further reducing port-related emissions.

From the perspective of the fuel categories discussed above, the classification of e-biofuels remains subject to debate. In terms of their production pathways, part of the organic carbon contained in e-biomass-to-liquid originates from conventional biomass-to-liquid processes and therefore falls under advanced biofuels. The remaining share of carbon is utilised through the addition of renewable hydrogen, and is thus categorised as RFNBOs. Consequently, in this study, the following additional policy-scenario constraints are applied.

\subsubsection*{Aviation sector constraints}

Let $D_{\text{aviation}}$ denote total aviation fuel demand (MWh). 
Let $E_{\text{bio}}^{(i)}$, $E_{\text{e-fuel}}^{(j)}$, and 
$E_{\text{e-bio}}^{(k)}$ represent the energy supplied by 
biofuels, e-fuels (RFNBO), and e-biofuels of type $i$, $j$, and $k$, respectively.

The overall renewable fuel requirement is formulated as

\begin{align}
\sum_i E_{\text{bio}}^{(i)} 
+ \sum_j E_{\text{e-fuel}}^{(j)} 
+ \sum_k E_{\text{e-bio}}^{(k)}
&\geq 0.70 \, D_{\text{aviation}}
\end{align}

To account for the RFNBO sub-target, only pure e-fuels and the hydrogen-derived fraction of e-biofuels are considered eligible. Using the carbon-efficiency-based allocation derived above, the hydrogen-derived share of e-biofuel $k$ is $\frac{\Delta \eta_c^{(k)}}{\eta_c^{\prime (k)}}$.

The RFNBO constraint therefore becomes

\begin{align}
\sum_j E_{\text{e-fuel}}^{(j)} 
+ \sum_k 
\left(
\frac{\Delta \eta_c^{(k)}}{\eta_c^{\prime (k)}}
E_{\text{e-bio}}^{(k)}
\right)
&\geq 0.35 \, D_{\text{aviation}}
\end{align}

\subsubsection*{Shipping sector constraint}

Let $D_{\text{shipping}}$ denote total marine fuel demand (MWh).

Given that shipping has not yet established specific RFNBO sub-targets, compliance is instead defined by an overall 80\% GHG-intensity reduction requirement. Considering that detailed GHG accounting involves upstream and downstream emissions as well as fuel- and gas-specific differences, a simplified assumption is adopted: renewable fuels must account for at least 80\% of total shipping energy demand. Accordingly,

\begin{align}
\sum_i E_{\text{bio}}^{(i)} 
+ \sum_j E_{\text{e-fuel}}^{(j)} 
+ \sum_k E_{\text{e-bio}}^{(k)}
&\geq 0.80 \, D_{\text{shipping}}
\end{align}

\subsection*{Key assumptions and scenarios}

Final energy demand in the aviation sector is assumed to increase by 70\% relative to current levels. The supply of aviation kerosene is determined endogenously within the energy system model.  
For the maritime sector, the fuel mix is exogenously constrained to 40\% methanol and 60\% oil-based fuels, reflecting a conservative transition pathway consistent with recent projections \cite{boyland2025methanol}.

Beyond the parameters varied in the sensitivity analyses, this study focuses on three structural determinants that critically shape long-term system outcomes:  
(i) the availability of geological carbon sequestration (CS),  
(ii) the inclusion or exclusion of e-biofuel technologies in the system optimisation, and  
(iii) the implementation of renewable fuel share mandates in aviation and shipping.

The EU targets a minimum \co\ sequestration capacity of 250~Mt~a$^{-1}$ by 2040 \cite{eu2024reg1735}, while assessments indicate a theoretical geological storage potential of around 507~Gt \co\ across Europe \cite{ec2015co2stop}. Given the substantial uncertainty surrounding the accessible share of this potential, we explore a wide range of CS availability levels: 140, 200, 250, 300, 350, 400, 600, 800, and 1000~Mt~\co~a$^{-1}$. This parametric variation allows us to assess how system costs, technology choices and fuel pathways depend on carbon storage availability and its interaction with e-biofuel deployment and policy constraints.

To capture these interactions, four integrated scenario configurations are constructed for each level of CS availability:

\begin{itemize}[noitemsep,leftmargin=*]
    \item \textbf{No-policy scenario:}  
    Aviation and maritime fuel supplies are optimised endogenously without renewable fuel share mandates.
    
    \item \textbf{Policy scenario:}  
    Aviation and maritime fuel supplies are optimised endogenously subject to renewable fuel share mandates.
    
    \item \textbf{Without e-biofuels:}  
    E-biofuel technologies are excluded from the technology portfolio.
    
    \item \textbf{With e-biofuels:}  
    E-biofuel technologies are available for investment and are optimised endogenously.
\end{itemize}

Combining the policy and technology dimensions yields four scenario cases:  
\textit{No policy + without e-biofuels},  
\textit{No policy + with e-biofuels},  
\textit{Policy + without e-biofuels}, and  
\textit{Policy + with e-biofuels}.

The analysis primarily focuses on comparing cost-effective energy system configurations with and without e-biofuels under the baseline no-policy scenario, while policy scenarios are used to assess the robustness of the results under renewable fuel mandates.


\subsection*{Limitations}

There are several limitations in this work. 
First, owing to computational constraints, the spatial resolution of the model is relatively coarse. While this may affect the absolute representation of infrastructure dependent outcomes, particularly in scenarios that distinguish the presence or absence of carbon dioxide and hydrogen transport networks, the comparative results across scenarios remain robust and consistently capture the underlying system trends. Second, aviation fuel demand in 2050 is specified exogenously using a growth factor that is subject to considerable uncertainty. Alternative demand trajectories would change the degree of resource scarcity, especially for biomass, renewable electricity and carbon sequestration, and thereby affect absolute deployment levels and system costs. While such changes could influence the magnitude of differences between scenarios, they are unlikely to alter the qualitative conclusions regarding the relative system value of different liquid fuel pathways. Third, the model does not account for the potential value of oxygen produced as a by product of electrolysis. In e-biofuel systems, hydrogen addition is accompanied by oxygen production, which could in principle be utilised in biomass gasification or upgrading processes and thereby reduce the need for dedicated oxygen supply \cite{bornemann2025oxygen}. Excluding this interaction may lead to a conservative assessment of the system level benefits of e-biofuels. Fourth, the model lacks detailed logistical modeling concerning the specific capacities (sizing) of biorefineries and the granular transport requirements for biomass collection and fuel distribution. Finally, while limited biomass imports are represented, the model does not consider international trade in liquid renewable fuels or liquid hydrogen. Imports of biofuels, e-fuels or e-biofuels could provide additional flexibility and reduce domestic resource constraints. At the same time, increased reliance on imported liquid fuels may expose the energy system to geopolitical risks \cite{desogus2023modelling}, external supply disruptions and strategic dependencies \cite{dejonghe2023natural}, particularly under conditions of globally rising demand for sustainable aviation fuels. By excluding liquid fuel imports, this analysis focuses on domestically deployable supply options and does not quantify trade-offs between cost efficiency and security of supply associated with higher import dependence.

\section*{Data availability}
The technology data used can be found at \href{https://github.com/yun-long-zhang/technology-data-electrobiofuel/tree/bio-efuel}{github.com/yun-long-zhang/technology-data-electrobiofuel/tree/bio-efuel} and is archived via Zenodo at: \href{https://zenodo.org/records/18985275}{https://zenodo.org/records/18985275} \cite{zeyen2026_technologydata_ebiofuel}.

Results files and code to generate figures are archived on Zenodo: \href{https://zenodo.org/records/18985101}{https://zenodo.org/records/18985101} \cite{zhang2026_ebiofuels_data}

\section*{Code availability}
The model code used can be found at \href{https://github.com/yun-long-zhang/pypsa-eur/tree/electrobiofuel}{github.com/yun-long-zhang/pypsa-eur/tree/electrobiofuel} and is archived via Zenodo at: \href{https://zenodo.org/records/18985259}{https://zenodo.org/records/18985259}  \cite{brown2026_pypsaeur}.

\section*{Acknowledgements}
We acknowledge funding from the Swedish Energy Agency, project numbers 2023-00888 (Y.Z. and M.M.), P2022-01082 (M.M) and P2023-01055 (F.H.). This research was partially funded by CETPartnership, the Clean Energy Transition Partnership under the 2022 joint call for research proposals, co-funded by the European Commission (grant agreement number 101069750) and with the funding organizations detailed on https://cetpartnership.eu/fundingagencies-and-call-modules.

The computations were enabled by resources provided by the National Academic Infrastructure for Supercomputing in Sweden (NAISS), partially funded by the Swedish Research Council through grant agreement no. 2022-06725.

\section*{Author Contributions Statement}
M.M. and F.H. conceived this study. M.M. designed the research together with F.H., Y.Z., K.P.; Y.Z. extended the model, collected the data and performed the models’ runs together with M.M., and with feedback from T.B.; Y.Z. performed the data analysis and created the visualisations, with the help from M.M. and F.H.; M.M., Y.Z., F.H., K.P., and T.B contributed to the interpretation of the results. M.M. and Y.Z. prepared the first draft. F.H., K.P., T.B. worked on the review and editing the paper. All authors approved and contributed to writing the paper.

\section*{Competing Interests Statement}
The authors declare no competing interests.

\bibliographystyle{elsarticle-num}
\bibliography{e_biofuels}

@article{Millinger2025,
  author    = {Millinger, M and Hedenus, F and Zeyen, E and Neumann, F and Reichenberg, L and Berndes, G},
  doi       = {10.1038/s41560-024-01693-6},
  issn      = {2058-7546},
  journal   = {Nature Energy},
  month     = {jan},
  number    = {February},
  pages     = {226--242},
  publisher = {Springer US},
  title     = {{Diversity of biomass usage pathways to achieve emissions targets in the European energy system}},
  url       = {https://www.nature.com/articles/s41560-024-01693-6},
  volume    = {10},
  year      = {2025}
}

@article{Bachorz2025,
  author    = {Bachorz, Clara and Verpoort, Philipp C. and Luderer, Gunnar and Ueckerdt, Falko},
  doi       = {10.1038/s41467-025-59277-1},
  issn      = {2041-1723},
  journal   = {Nature Communications},
  month     = {apr},
  number    = {1},
  pages     = {3984},
  publisher = {Springer US},
  title     = {{Exploring techno-economic landscapes of abatement options for hard-to-electrify sectors}},
  url       = {https://www.nature.com/articles/s41467-025-59277-1},
  volume    = {16},
  year      = {2025}
}

@article{EU2023_aviation,
  author  = {{European Union}},
  journal = {Official Journal of the European Union},
  number  = {401},
  pages   = {1--30},
  title   = {{Regulation (EU) 2023/2405 of the European Parliament and of the Council of 18 October 2023 on ensuring a level playing field for sustainable air transport (ReFuelEU Aviation)}},
  url     = {http://data.europa.eu/eli/reg/2023/2405/oj},
  volume  = {2405},
  year    = {2023}
}

@article{EU2023_maritime,
  author  = {{European Union}},
  journal = {Official Journal of the European Union},
  number  = {July},
  pages   = {48--100},
  title   = {{Regulation (EU) 2023/1805 of the European Parliament and of the Council of 13 September 2023 on the use of renewable and low-carbon fuels in maritime transport, and amending Directive 2009/16/EC (Text with EEA relevance)}},
  url     = {http://data.europa.eu/eli/reg/2023/1805/oj},
  volume  = {66},
  year    = {2023}
}

@article{Mesfun2023,
  author   = {Mesfun, Sennai and Gustafsson, Gabriel and Larsson, Anton and Samavati, Mahrokh},
  doi      = {10.3390/en16217436},
  journal  = {Energies},
  number   = {21},
  pages    = {7436},
  title    = {{Electrification of Biorefinery Concepts for Improved Productivity --- Yield, Economic and GHG Performances}},
  volume   = {16},
  year     = {2023}
}

@article{Cherp2021,
  author    = {Cherp, Aleh and Vinichenko, Vadim and Tosun, Jale and Gordon, Joel A. and Jewell, Jessica},
  doi       = {10.1038/s41560-021-00863-0},
  issn      = {20587546},
  journal   = {Nature Energy},
  number    = {7},
  pages     = {742--754},
  publisher = {Springer US},
  title     = {{National growth dynamics of wind and solar power compared to the growth required for global climate targets}},
  volume    = {6},
  year      = {2021}
}

@article{Allen2022,
  author    = {Allen, Myles R and Friedlingstein, Pierre and Girardin, C{\'e}cile AJ and Jenkins, Stuart and Malhi, Yadvinder and Mitchell-Larson, Eli and Peters, Glen P and Rajamani, Lavanya},
  doi       = {10.1146/annurev-environ-112320-105050},
  journal   = {Annual Review of Environment and Resources},
  volume    = {47},
  number    = {1},
  pages     = {849--887},
  year      = {2022},
  publisher = {Annual Reviews},
  title     = {{Net zero: science, origins, and implications}}
}

@article{Kazlou2024,
  author    = {Kazlou, Tsimafei and Cherp, Aleh and Jewell, Jessica},
  doi       = {10.1038/s41558-024-02104-0},
  issn      = {1758-678X},
  journal   = {Nature Climate Change},
  month     = {sep},
  volume    = {14},
  number    = {10},
  pages     = {1047--1055},
  title     = {{Feasible deployment of carbon capture and storage and the requirements of climate targets}},
  year      = {2024}
}

@article{Grant2022,
  author   = {Grant, Neil and Gambhir, Ajay and Mittal, Shivika and Greig, Chris and K{\"{o}}berle, Alexandre C.},
  doi      = {10.1016/j.ijggc.2022.103766},
  issn     = {17505836},
  journal  = {International Journal of Greenhouse Gas Control},
  number   = {August},
  title    = {{Enhancing the realism of decarbonisation scenarios with practicable regional constraints on CO2 storage capacity}},
  volume   = {120},
  year     = {2022}
}

@article{Gidden2025,
  author    = {Gidden, Matthew and Joshi, Siddharth and Armitage, John and Christ, Alina-Berenice and Boettcher, Miranda and Brutschin, Elina and Koberle, Alex and Schellnhuber, Hans Joachim and Schleussner, Carl-Friedrich and Riahi, Keywan and Rogelj, Joeri},
  doi       = {10.1038/s41586-025-09423-y},
  issn      = {1476-4687},
  journal   = {Nature},
  number    = {September},
  publisher = {Springer US},
  title     = {{A prudent planetary boundary for geological carbon storage}},
  volume    = {645},
  year      = {2025}
}

@article{Dimanchev2025,
  author  = {Dimanchev, Emil and Gabriel, Steven and Fleten, Stein-Erik and Pecci, Filippo and Korp{\aa}s, Magnus},
  journal = {Available at SSRN 5636369},
  title   = {{Choosing climate policies in a second-best world with incomplete risk markets}},
  year    = {2025}
}

@article{Ueckerdt2021,
  author  = {Ueckerdt, Falko and Bauer, Christian and Dirnaichner, Alois and Everall, Jordan and Sacchi, Romain and Luderer, Gunnar},
  doi     = {10.1038/s41558-021-01032-7},
  journal = {Nature Climate Change},
  volume  = {11},
  pages   = {384--393},
  title   = {{Potential and risks of hydrogen-based e-fuels in climate change mitigation}},
  year    = {2021}
}

@software{glaum2025minMeOH,
  author  = {Glaum, Philipp and Neumann, Fabian and Millinger, Markus and Brown, Tom},
  title   = {min-MeOH-economy: Minimal methanol economy code repository},
  version = {master},
  year    = {2025},
  url     = {https://github.com/p-glaum/min-MeOH-economy/tree/master},
  note    = {GitHub repository, accessed 2025-11-21}
}

@techreport{ec2015co2stop,
  author      = {{European Commission}},
  title       = {{CO2StoP -- Assessment of the CO2 Storage Potential in Europe}},
  year        = {2015},
  institution = {European Commission},
  address     = {Brussels, Belgium},
  note        = {Final Report},
  url         = {https://data.europa.eu/doi/10.2777/91746}
}

@misc{eu2024reg1735,
  author       = {{European Commission}},
  title        = {{Regulation (EU) 2024/1735 of the European Parliament and of the Council of 13 June 2024 on establishing a framework of measures for strengthening Europe's net-zero technology manufacturing ecosystem and amending Regulation (EU) 2018/1724}},
  year         = {2024},
  howpublished = {Official Journal of the European Union, L 1735, 28 June 2024},
  note         = {Text with EEA relevance. PE/45/2024/REV/1. Current consolidated version as of 17 August 2025.},
  url          = {http://data.europa.eu/eli/reg/2024/1735/oj},
  institution  = {European Parliament and the Council of the European Union},
  language     = {English}
}

@dataset{jrc_enspreso2021,
  author       = {{European Commission, Joint Research Centre (JRC)}},
  year         = {2021},
  title        = {{ENSPRESO -- INTEGRATED DATA}},
  institution  = {European Commission, Joint Research Centre (JRC)},
  type         = {Dataset},
  howpublished = {\url{http://data.europa.eu/89h/88d1c405-0448-4c9e-b565-3c30c9b167f7}},
  note         = {PID: http://data.europa.eu/89h/88d1c405-0448-4c9e-b565-3c30c9b167f7}
}

@article{brown2018synergies,
  author    = {Brown, Tom and Schlachtberger, David and Kies, Alexander and Schramm, Stefan and Greiner, Martin},
  doi       = {10.1016/j.energy.2018.06.222},
  journal   = {Energy},
  volume    = {160},
  pages     = {720--739},
  year      = {2018},
  publisher = {Elsevier},
  title     = {{Synergies of sector coupling and transmission reinforcement in a cost-optimised, highly renewable European energy system}}
}

@article{neumann2023potential,
  author    = {Neumann, Fabian and Zeyen, Elisabeth and Victoria, Marta and Brown, Tom},
  doi       = {10.1016/j.joule.2023.06.016},
  journal   = {Joule},
  volume    = {7},
  number    = {8},
  pages     = {1793--1817},
  year      = {2023},
  publisher = {Elsevier},
  title     = {{The potential role of a hydrogen network in Europe}}
}

@article{neumann2022assessments,
  author    = {Neumann, Fabian and Hagenmeyer, Veit and Brown, Tom},
  doi       = {10.1016/j.apenergy.2022.118859},
  journal   = {Applied Energy},
  volume    = {314},
  pages     = {118859},
  year      = {2022},
  publisher = {Elsevier},
  title     = {{Assessments of linear power flow and transmission loss approximations in coordinated capacity expansion problems}}
}

@article{mantzos2017jrc,
  author  = {Mantzos, L and Wiesenthal, T and Matei, NA and Tchung-Ming, S and Rozsai, M and Russ, HP and Ramirez, A Soria},
  doi     = {10.2760/182725},
  journal = {Methodological Note},
  title   = {{JRC-IDEES: Integrated Database of the European Energy Sector}},
  year    = {2017}
}

@article{zeyen2021mitigating,
  author    = {Zeyen, Elisabeth and Hagenmeyer, Veit and Brown, Tom},
  doi       = {10.1016/j.energy.2021.120784},
  journal   = {Energy},
  volume    = {231},
  pages     = {120784},
  year      = {2021},
  publisher = {Elsevier},
  title     = {{Mitigating heat demand peaks in buildings in a highly renewable European energy system}}
}

@article{victoria2022speed,
  author    = {Victoria, Marta and Zeyen, Elisabeth and Brown, Tom},
  doi       = {10.1016/j.joule.2022.04.016},
  journal   = {Joule},
  volume    = {6},
  number    = {5},
  pages     = {1066--1086},
  year      = {2022},
  publisher = {Elsevier},
  title     = {{Speed of technological transformations required in Europe to achieve different climate goals}}
}

@article{millinger2022biofuel,
  author    = {Millinger, Markus and Reichenberg, Lina and Hedenus, Fredrik and Berndes, G{\"o}ran and Zeyen, Elisabeth and Brown, Tom},
  doi       = {10.1016/j.apenergy.2022.120016},
  journal   = {Applied Energy},
  volume    = {326},
  pages     = {120016},
  year      = {2022},
  publisher = {Elsevier},
  title     = {{Are biofuel mandates cost-effective? -- An analysis of transport fuels and biomass usage to achieve emissions targets in the European energy system}}
}

@techreport{boyland2025methanol,
  author      = {Boyland, Joe and Annisa, Aninda and Kirketerp-M{\o}ller, Trine and Cremer, Justin},
  title       = {{From Pilots to Practice: Methanol and Ammonia as Shipping Fuels}},
  institution = {Global Maritime Forum for the Getting to Zero Coalition},
  year        = {2025},
  month       = {August},
  note        = {This report builds upon the \textit{Mapping of zero-emission pilots and demonstration projects} series first published in 2020.},
  url         = {https://www.globalmaritimeforum.org/},
  address     = {Copenhagen, Denmark}
}

@article{ruiz2019enspreso,
  author    = {Ruiz, Pablo and Nijs, Wouter and Tarvydas, Dalius and Sgobbi, Alessandra and Zucker, Andreas and Pilli, Roberto and Jonsson, Ragnar and Camia, Andrea and Thiel, Christine and Hoyer-Klick, Carsten and others},
  doi       = {10.1016/j.esr.2019.100379},
  journal   = {Energy Strategy Reviews},
  volume    = {26},
  pages     = {100379},
  year      = {2019},
  publisher = {Elsevier},
  title     = {{ENSPRESO -- an open, {EU}-28 wide, transparent and coherent database of wind, solar and biomass energy potentials}}
}

@article{Odenweller2022,
  author    = {Odenweller, Adrian and Ueckerdt, Falko and Nemet, Gregory F and Jensterle, Miha and Luderer, Gunnar},
  doi       = {10.1038/s41560-022-01097-4},
  issn      = {2058-7546},
  journal   = {Nature Energy},
  month     = {sep},
  publisher = {Springer US},
  title     = {{Probabilistic feasibility space of scaling up green hydrogen supply}},
  year      = {2022}
}

@article{hannula2016hydrogen,
  author    = {Hannula, Ilkka},
  doi       = {10.1016/j.energy.2016.03.119},
  journal   = {Energy},
  volume    = {104},
  pages     = {199--212},
  year      = {2016},
  publisher = {Elsevier},
  title     = {{Hydrogen enhancement potential of synthetic biofuels manufacture in the European context: A techno-economic assessment}}
}

@article{anetjarvi2023benefits,
  author    = {Anetj{\"a}rvi, Eemeli and Vakkilainen, Esa and Melin, Kristian},
  doi       = {10.1016/j.energy.2023.127202},
  journal   = {Energy},
  volume    = {276},
  pages     = {127202},
  year      = {2023},
  publisher = {Elsevier},
  title     = {{Benefits of hybrid production of e-methanol in connection with biomass gasification}}
}

@techreport{furusjo2022bioelectrofuels,
  author      = {Furusj{\"o}, Erik and Mesfun, Sennai and Samavati, Mahrokh and Larsson, Anton and Gustafsson, Gabriel
                 and Gustavsson, Christer and Hermansson, Sven and Olsson, Johanna and Sundell, Per
                 and Kairento, Kajsa and Karlsson, Per-Arne and Werner, Linda and Ahlstr{\"o}m, Johan
                 and Hamon, Camille and Gunneberg, Tobias},
  title       = {{Bio-electro fuels: Hybrid technology for improved resource efficiency}},
  institution = {RISE Research Institutes of Sweden},
  year        = {2022},
  type        = {Report},
  number      = {FDOS 45:2022},
  url         = {https://f3centre.se/en/renewable-transportation-fuels-and-systems/},
  note        = {Published within the f3 -- Swedish Knowledge Centre for Renewable Transportation Fuels}
}

@article{hillestad2018improving,
  author    = {Hillestad, Magne and Ostadi, Mohammad and Serrano, Gd Alamo and Rytter, Erling and Austb{\o}, Bj{\o}rn and Pharoah, JG and Burheim, Odne Stokke},
  doi       = {10.1016/j.fuel.2018.08.004},
  journal   = {Fuel},
  volume    = {234},
  pages     = {1431--1451},
  year      = {2018},
  publisher = {Elsevier},
  title     = {{Improving carbon efficiency and profitability of the biomass to liquid process with hydrogen from renewable power}}
}

@article{grahn2022review,
  author    = {Grahn, Maria and Malmgren, Elin and Korberg, Andrei D and Taljegard, Maria and Anderson, James E and Brynolf, Selma and Hansson, Julia and Skov, Iva Ridjan and Wallington, Timothy J},
  doi       = {10.1088/2516-1083/ac7571},
  journal   = {Progress in Energy},
  volume    = {4},
  number    = {3},
  pages     = {032010},
  year      = {2022},
  publisher = {IOP Publishing},
  title     = {{Review of electrofuel feasibility --- cost and environmental impact}}
}

@article{Korberg2021b,
author = {Korberg, A D and Brynolf, S and Grahn, M and Skov, I R},
doi = {10.1016/j.rser.2021.110861},
issn = {1364-0321},
journal = {Renewable and Sustainable Energy Reviews},
number = {March},
pages = {110861},
publisher = {Elsevier Ltd},
title = {{Techno-economic assessment of advanced fuels and propulsion systems in future fossil-free ships}},
url = {https://doi.org/10.1016/j.rser.2021.110861},
volume = {142},
year = {2021}
}

@misc{EU2024_3012,
  author       = {{European Commission}},
  title        = {{Regulation (EU) 2024/3012 of the European Parliament and of the Council of 27 November 2024 establishing a Union certification framework for permanent carbon removals, carbon farming and carbon storage in products}},
  year         = {2024},
  howpublished = {Official Journal of the European Union, L 2024/3012, 6 December 2024},
  institution  = {European Parliament and Council of the European Union},
  note         = {PE/92/2024/REV/1. In force. Text with EEA relevance},
  url          = {http://data.europa.eu/eli/reg/2024/3012/oj}
}

@article{ostadi2019boosting,
  author    = {Ostadi, Mohammad and Rytter, Erling and Hillestad, Magne},
  doi       = {10.1016/j.biombioe.2019.105282},
  journal   = {Biomass and Bioenergy},
  volume    = {127},
  pages     = {105282},
  year      = {2019},
  publisher = {Elsevier},
  title     = {{Boosting carbon efficiency of the biomass to liquid process with hydrogen from power: The effect of H2/CO ratio to the Fischer-Tropsch reactors on the production and power consumption}}
}

@article{katla2024synthetic,
  author    = {Katla-Milewska, Daria and Nazir, Shareq Mohd and Skorek-Osikowska, Anna},
  doi       = {10.1016/j.enconman.2023.117895},
  journal   = {Energy Conversion and Management},
  volume    = {300},
  pages     = {117895},
  year      = {2024},
  publisher = {Elsevier},
  title     = {{Synthetic natural gas (SNG) production with higher carbon recovery from biomass: Techno-economic assessment}}
}

@article{park2025bio,
  author    = {Park, Jiye and Qi, Meng and Baek, Jaeho and Choi, Dongho and Kwon, Eilhann E and Cho, Hyungtae and Lee, Jaewon},
  doi       = {10.1016/j.enconman.2025.120052},
  journal   = {Energy Conversion and Management},
  volume    = {341},
  pages     = {120052},
  year      = {2025},
  publisher = {Elsevier},
  title     = {{Bio-e-methanol production via biogas partial oxidation integrated with solid oxide electrolyzer cell: A comprehensive energy, exergy, economic, and environmental (4E) analysis}}
}

@article{kumar2025advancements,
  author    = {Kumar, Sonu and Palange, Rupesh and De Blasio, Cataldo},
  doi       = {10.1039/D5SE00504C},
  journal   = {Sustainable Energy \& Fuels},
  volume    = {9},
  number    = {18},
  pages     = {4793--4831},
  year      = {2025},
  publisher = {Royal Society of Chemistry},
  title     = {{Advancements in gasification technologies: insights into modeling studies, power-to-X applications and sustainability assessments}}
}

@article{patel2025review,
  author    = {Patel, Ronak and Rajaraman, TS and Rana, Paresh H and Ambegaonkar, Nikita J and Patel, Sanjay},
  doi       = {10.1016/j.rechem.2025.102052},
  journal   = {Results in Chemistry},
  volume    = {13},
  pages     = {102052},
  year      = {2025},
  publisher = {Elsevier},
  title     = {{A review on techno-economic analysis of lignocellulosic biorefinery producing biofuels and high-value products}}
}

@article{martulli2025potential,
  author    = {Martulli, Alessandro and Brandt, Kristin and Allroggen, Florian and Malina, Robert},
  doi       = {10.1038/s41467-025-58955-2},
  journal   = {Nature Communications},
  year      = {2025},
  publisher = {Nature Publishing Group UK London},
  title     = {{The potential scale-up of sustainable aviation fuels production capacity to meet global and {EU} policy targets}}
}

@misc{genasolutions2025,
  author       = {{GENA Solutions}},
  title        = {{Analysis and Insights: 57}},
  year         = {2025},
  howpublished = {\url{https://www.genasolutions.com/analysis_and_insights/57}},
  note         = {Accessed: 2026-01-02}
}

@article{Mignone2024,
author = {Mignone, Bryan K. and Clarke, Leon and Edmonds, James A. and Gurgel, Angelo and Herzog, Howard J. and Johnson, Jeremiah X. and Mallapragada, Dharik S. and McJeon, Haewon and Morris, Jennifer and O'Rourke, Patrick R. and Paltsev, Sergey and Rose, Steven K. and Steinberg, Daniel C. and Venkatesh, Aranya},
doi = {10.1038/s41467-024-47059-0},
issn = {20411723},
journal = {Nature Communications},
number = {1},
pmid = {38729928},
publisher = {Springer US},
title = {{Drivers and implications of alternative routes to fuels decarbonization in net-zero energy systems}},
volume = {15},
year = {2024}
}

@misc{eu2025climate_law_amendment,
  author  = {{European Commission}},
  title   = {{Proposal for a Regulation of the European Parliament and of the Council amending Regulation (EU) 2021/1119 establishing the framework for achieving climate neutrality}},
  year    = {2025},
  number  = {COM(2025) 524 final},
  address = {Brussels},
  note    = {Legislative proposal},
  url     = {https://eur-lex.europa.eu/legal-content/EN/TXT/?uri=celex:52025PC0524}
}

@misc{EC_REPowerEU_2022,
  author       = {{European Commission}},
  title        = {{REPowerEU Plan}},
  howpublished = {Communication from the Commission},
  number       = {COM(2022) 230 final},
  year         = {2022},
  month        = may,
  institution  = {European Commission},
  url          = {https://eur-lex.europa.eu/legal-content/EN/TXT/?uri=COM:2022:230:FIN}
}

@article{bube2024power,
  author    = {Bube, Stefan and Sens, Lucas and Drawer, Chris and Kaltschmitt, Martin},
  doi       = {10.1016/j.enconman.2024.118220},
  journal   = {Energy Conversion and Management},
  volume    = {304},
  pages     = {118220},
  year      = {2024},
  publisher = {Elsevier},
  title     = {{Power and biogas to methanol -- a techno-economic analysis of carbon-maximized green methanol production via two reforming approaches}}
}

@article{beiron2026locked,
  author    = {Beiron, Johanna and Harvey, Simon and Thunman, Henrik},
  doi       = {10.1016/j.fuel.2025.137181},
  journal   = {Fuel},
  volume    = {406},
  pages     = {137181},
  year      = {2026},
  publisher = {Elsevier},
  title     = {{Locked in on RFNBOs -- Will {EU} mandates for drop-in synthetic aviation fuels lead to decreased energy- and cost-efficiency?}}
}

@techreport{Pettersson_RISE_REDFuels_2025,
  author      = {Pettersson, Karin and Wickman, Clara},
  title       = {{Classification and sustainability criteria for renewable fuels in the {EU} -- what actually applies?}},
  institution = {Research Institutes of Sweden (RISE)},
  year        = {2025},
  number      = {P2023-00841},
  address     = {Gothenburg, Sweden},
  note        = {Final project report commissioned by the Swedish Energy Agency (Energimyndigheten)},
  url         = {https://www.ri.se/sites/default/files/2025-10/Slutrapport_Klassificering%20och%20h%C3%A5llbarhetskriterier%20f%C3%B6r%20f%C3%B6rnybara%20drivmedel%20i%20EU.pdf}
}

@article{bornemann2025oxygen,
  author    = {Bornemann, Luka and Lange, Jelto and Kaltschmitt, Martin},
  doi       = {10.1016/j.enconman.2025.120213},
  journal   = {Energy Conversion and Management},
  volume    = {344},
  pages     = {120213},
  year      = {2025},
  publisher = {Elsevier},
  title     = {{Oxygen production via electrolysis: A model-based assessment of its impact on a climate-neutral German energy system}}
}

@article{desogus2023modelling,
  author    = {Desogus, Eleonora and Grosso, Daniele and Bompard, Ettore and Russo, Stefano Lo},
  doi       = {10.1016/j.energy.2023.128578},
  journal   = {Energy},
  volume    = {284},
  pages     = {128578},
  year      = {2023},
  publisher = {Elsevier},
  title     = {{Modelling the geopolitical impact on risk assessment of energy supply system: The case of Italian crude oil supply}}
}

@article{dejonghe2023natural,
  author    = {Dejonghe, Marie and Van de Graaf, Thijs and Belmans, Ronnie},
  doi       = {10.1016/j.erss.2023.103301},
  journal   = {Energy Research \& Social Science},
  volume    = {106},
  pages     = {103301},
  year      = {2023},
  publisher = {Elsevier},
  title     = {{From natural gas to hydrogen: Navigating import risks and dependencies in Northwest Europe}}
}

@misc{EU_DelegatedReg_2023_1185,
  author       = {{European Commission}},
  title        = {{Commission Delegated Regulation (EU) 2023/1185 of 10 February 2023 supplementing Directive (EU) 2018/2001 by establishing a minimum threshold for greenhouse gas emissions savings of recycled carbon fuels and specifying a methodology for assessing greenhouse gas emissions savings from RFNBOs and recycled carbon fuels}},
  year         = {2023},
  note         = {OJ L 157, 20.6.2023, pp. 20--33},
  howpublished = {Official Journal of the European Union},
  url          = {http://data.europa.eu/eli/reg_del/2023/1185/oj}
}

@techreport{EC_HydrogenDelegatedActs_QA_2024,
  author      = {{European Commission}},
  title       = {{Q\&A on the Implementation of Hydrogen Delegated Acts}},
  institution = {European Commission},
  year        = {2024},
  number      = {Version 1.0},
  note        = {Draft guidance document},
  url         = {https://circabc.europa.eu/ui/group/8f5f9424-a7ef-4dbf-b914-1af1d12ff5d2/library/ca8efd4d-cb44-4aec-914d-3d95f95ea293/details}
}

@article{schreyer2025net,
  author    = {Schreyer, Felix and Ueckerdt, Falko and Pietzcker, Robert and Odenweller, Adrian and Merfort, Anne and Rodrigues, Renato and Strefler, Jessica and L{\'e}cuyer, Fabrice and Luderer, Gunnar},
  doi       = {10.1038/s41467-025-65930-2},
  journal   = {Nature Communications},
  volume    = {16},
  number    = {1},
  pages     = {10700},
  year      = {2025},
  publisher = {Nature Publishing Group UK London},
  title     = {{From net-zero to zero-fossil in transforming the {EU} energy system}}
}

@article{de2026biofuels,
  author    = {de Chambost, Etienne and Merceron, Louis and Boissonnet, Guillaume},
  journal   = {Sustainable Energy \& Fuels},
  year      = {2026},
  publisher = {Royal Society of Chemistry},
  title     = {{From biofuels to e-fuels: an assessment of techno-economic and environmental performance}},
  doi       = {10.1039/D5SE00786K}
}

@dataset{zhang2026_ebiofuels_data,
  author    = {Zhang, Yunlong},
  title     = {{Results analysis of e-biofuels study}},
  year      = {2026},
  publisher = {Zenodo},
  version   = {v1.0},
  doi       = {10.5281/zenodo.18985101},
  url       = {https://doi.org/10.5281/zenodo.18985101}
}

@software{brown2026_pypsaeur,
  author    = {Brown, Tom and Victoria, Marta and Zeyen, Emma and Hofmann, Fabian and Neumann, Fabian and Frysztacki, Micha{\l} and Hampp, Johannes and Schlachtberger, David and H{\"o}rsch, Jonas and Schledorn, Amos and Schau{\ss}, Christoph and van Greevenbroek, Koen and Millinger, Markus and Glaum, Philipp and Xiong, Bo and Seibold, Tobias},
  title     = {{PyPSA-Eur: An open sector-coupled optimisation model of the European energy system}},
  year      = {2026},
  version   = {v1.0},
  publisher = {Zenodo},
  doi       = {10.5281/zenodo.18985259},
  url       = {https://doi.org/10.5281/zenodo.18985259}
}

@dataset{zeyen2026_technologydata_ebiofuel,
  author    = {Zeyen, Lisa and Hampp, Johannes and Neumann, Fabian and Parzen, Max and Alamian, Alireza and Franken, Lukas and Brown, Tom and Geis, Julian and Glaum, Philipp and Finozzi, Fabrizio and van der Plas, Marta and Millinger, Markus and Gilon, Thomas and Lerede, Daniele and Schledorn, Amos and Trippe, Lukas and others},
  title     = {{Technology-data-electrobiofuel: e-biofuels dataset}},
  year      = {2026},
  version   = {v1.0},
  publisher = {Zenodo},
  doi       = {10.5281/zenodo.18985275},
  url       = {https://doi.org/10.5281/zenodo.18985275}
}

\clearpage
\onecolumn
\appendix

\setcounter{figure}{0}
\setcounter{table}{0}
\setcounter{equation}{0}

\renewcommand{\thefigure}{S\arabic{figure}}
\renewcommand{\thetable}{S\arabic{table}}
\renewcommand{\theequation}{S\arabic{equation}}

\section*{Supplementary Information}
\setcounter{page}{1}
\noindent\textbf{Contents}

\vspace{0.5em}

\noindent\textbf{Supplementary Figures}
\begin{itemize}
    \item Figure S1: Liquid fuel mix in the European energy system
    \item Figure S2: CO$_2$ and solid biomass shadow prices
    \item Figure S3: Power generation, hydrogen production, biomass consumption and CO$_2$ flows
    \item Figure S4: Total liquid fuel production cost differences
    \item Figure S5: Merit order of aviation fuels under different CO$_2$ sequestration levels
    \item Figure S6: System cost and fuel mix under policy scenarios
    \item Figure S7: Deployment shares under varying electrolyser, biomass and CO$_2$ constraints
\end{itemize}

\vspace{0.5em}

\noindent\textbf{Supplementary Tables}
\begin{itemize}
    \item Table S1: Definitions and classification of liquid fuel technologies considered in this study
\end{itemize}

\clearpage

\section*{Supplementary Figures}
\setcounter{figure}{0}
\begin{figure}[!htbp]
\centering
\includegraphics[width=\linewidth]{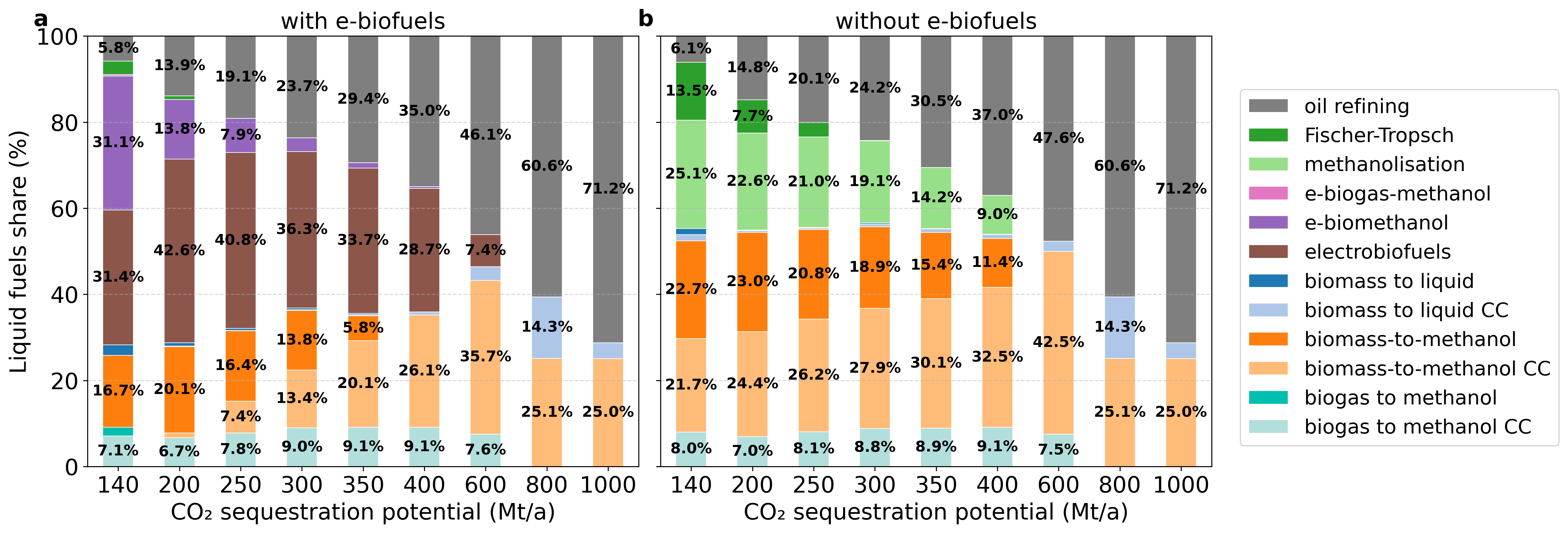}
\caption{Liquid fuel mix in the European energy system. Panels (a) and (b) show the shares of different liquid fuel categories under the baseline scenario \textit{with} and \textit{without} e-biofuels, respectively.}
\label{fig:S1_detail_fuel_mix}
\end{figure}

\clearpage
\begin{figure}[p]
\centering
\includegraphics[width=\linewidth]{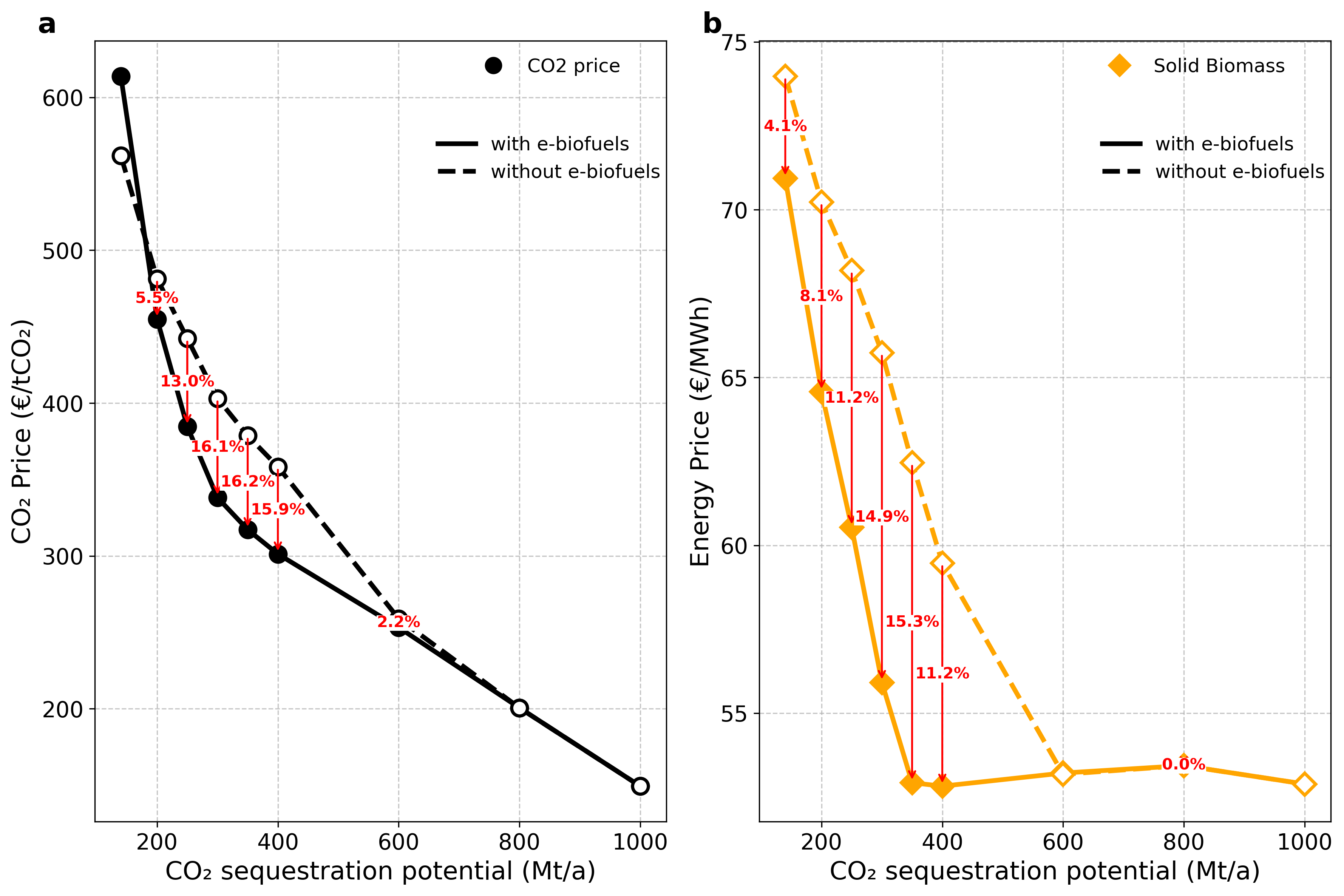}
\caption{CO$_2$ (a) and solid biomass (b) shadow price comparison under the baseline scenario.}
\label{fig:S2_CO2_Price}
\end{figure}

\clearpage
\begin{figure}[p]
\centering
\includegraphics[width=\linewidth]{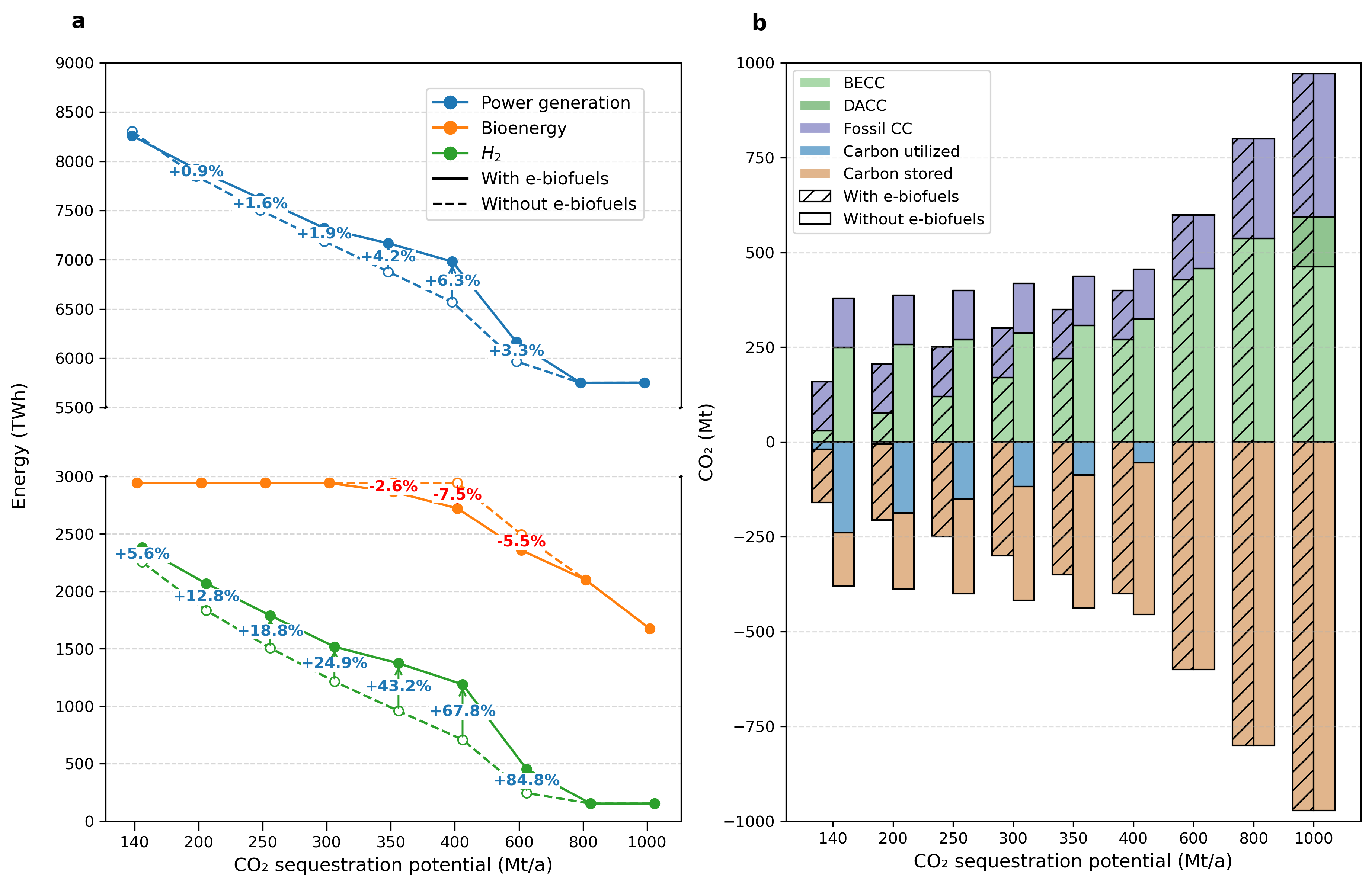}
\caption{Power generation, H$_2$ production, solid biomass resources consumption and CO$_2$ flow in different scenarios. BECC: bioenergy with carbon capture, and DACC: direct air carbon capture.}
\label{fig:S3_el_bio_CC_sweep}
\end{figure}

\clearpage
\begin{figure}[p]
\centering
\includegraphics[width=\linewidth]{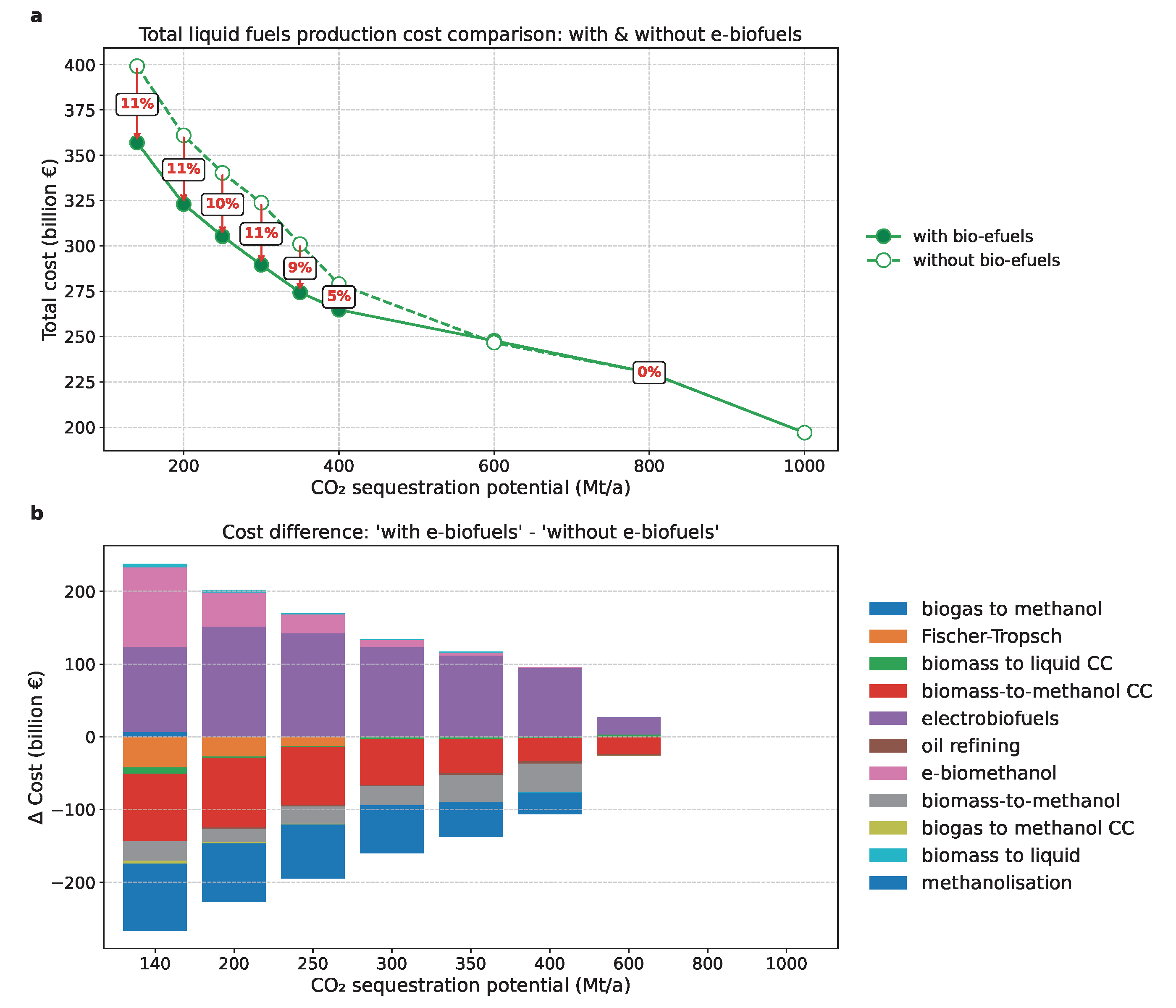}
\caption{Total liquid fuels production cost difference under the baseline scenario. Panels (a) shows the total cost comparison and panels (b) shows the main cost difference sources.}
\label{fig:S3_liquid_total_cost_nopolicy}
\end{figure}

\clearpage
\begin{figure}[p]
\centering
\includegraphics[width=\linewidth]{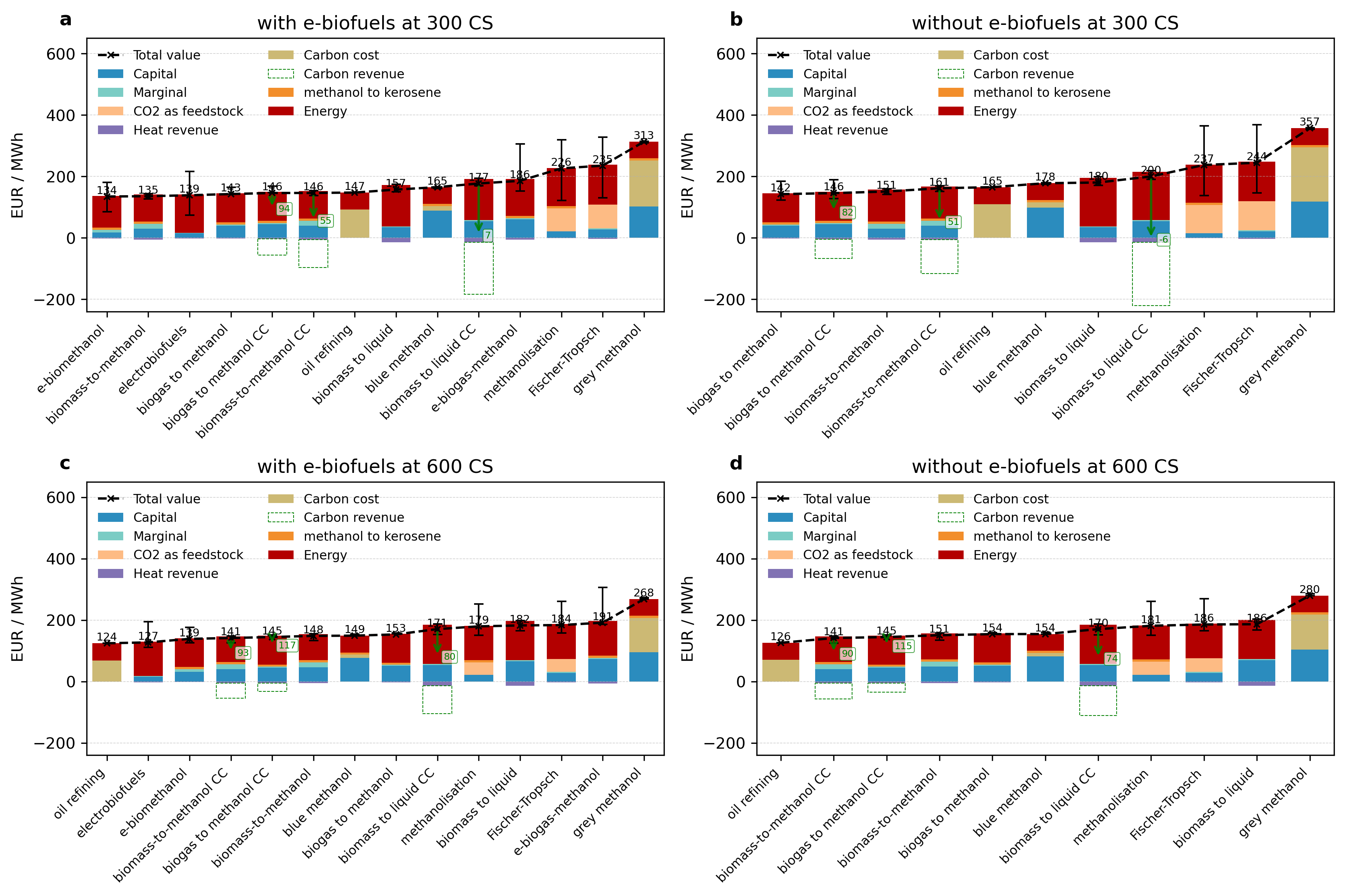}
\caption{Merit order of liquid fuels using in aviation under baseline scenarios with carbon sequestration (CS) potentials of 300~Mt~a$^{-1}$ and 600~Mt~a$^{-1}$. Panels (a) and (b) show the merit order for 300~Mt~a$^{-1}$ \textit{with} and \textit{without} e-biofuels, respectively. Panels (c) and (d) show the merit order for 600~Mt~a$^{-1}$ \textit{with} and \textit{without} e-biofuels, respectively.}
\label{fig:S4_merit_order_300_600CS}
\end{figure}

\clearpage
\begin{figure}[p]
\centering
\includegraphics[width=\linewidth]{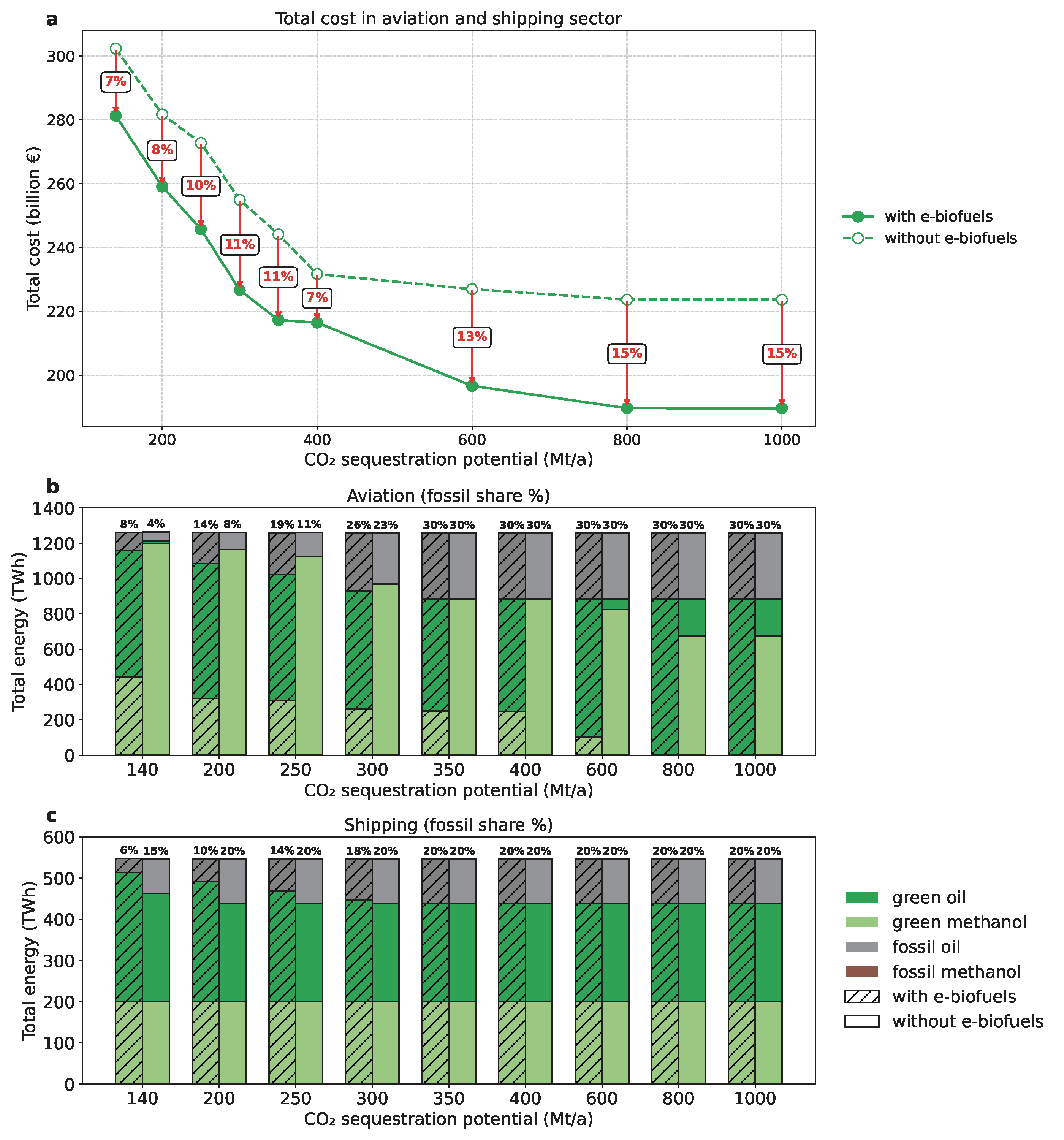}
\caption{Total system cost (a) and fuel mix in the aviation (b) and shipping (c) sectors under policy scenarios.}
\label{fig:S5_two_sector_cost}
\end{figure}

\begin{figure}[!htp]
    \centering
    \includegraphics[width=1\linewidth]{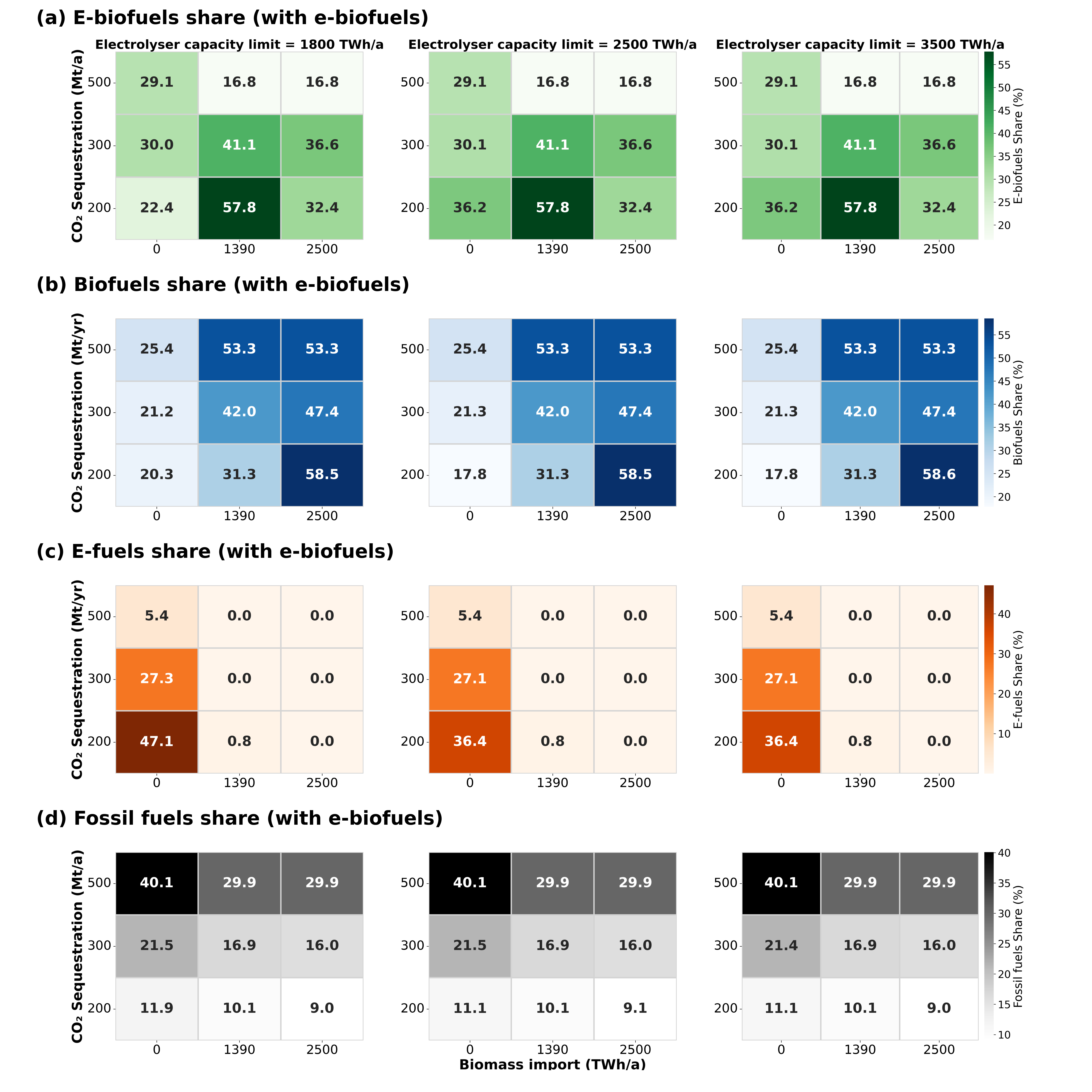}
    \caption{\textbf{Deployment shares of different liquid fuel pathways under varying \co\ electrolyser capacity limits and biomass import scenarios.} The analysis considers three electrolyser production limits (1800, 2500, and 3500 TWh/a), three biomass import levels (0, 1390, and 2500 TWh/a, where 1390 TWh/a represents 5 EJ/a), and three CO$_2$ sequestration potentials (200, 300, and 500 Mt/a). Panels show the fuel shares in total liquid fuel production when e-biofuel technologies are available, for (a) E-biofuels, (b) Biofuels, (c) E-fuels, and (d) Fossil fuels. Across the nine sensitivity cases (three biomass-import levels $\times$ three CO$_2$ sequestration levels), the dominant liquid-fuel option follows a clear and structured pattern. Biofuels dominate only under the most biomass- and CCS-abundant conditions, i.e., when biomass imports substantially exceed domestic availability ($\approx 2\times$) and CO$_2$ sequestration potential is very high (approaching $\approx 2\times$ the EU’s 2040 sequestration target). By contrast, e-fuels dominate in biomass-scarce, low-CCS settings, consistent with greater reliance on electricity-based synthesis when sustainable biomass is limited and carbon storage is constrained. Fossil-derived fuels become dominant only at the high-CCS end of the range, whereas e-biofuels dominate under more intermediate conditions, with biomass imports comparable to domestic availability and CO$_2$ sequestration potential not exceeding the EU 2040 target. Overall, the results suggest that biofuels, e-fuels and fossil fuels expand primarily in “corner” (more extreme) scenarios, while e-biofuels gain the largest share in more moderate settings, arguably indicating a comparatively lower-exposure pathway under plausible real-world constraints.}
    \label{fig:biomass_CS_h2_sensitivity}
\end{figure}

\clearpage

\section*{Supplementary Tables}
\setcounter{table}{0}

\begin{table}[H]
\centering
\small
\begin{tabular}{p{4.8cm} p{9.2cm} p{3.2cm}}
\toprule
\textbf{Technology} & \textbf{Description} & \textbf{Category} \\
\midrule
Biomass to liquid (BtL) 
& Conversion of solid biomass into liquid hydrocarbon fuels via thermochemical gasification and subsequent synthesis (e.g., Fischer--Tropsch), without carbon capture. 
& Biofuels \\

Biomass to liquid with carbon capture (BtL CC) 
& Biomass-to-liquid fuel production combined with capture of excess CO$_2$ from the conversion process. 
& Biofuels \\

Biomass-to-methanol 
& Production of methanol from biomass-derived syngas without carbon capture. 
& Biofuels \\

Biomass-to-methanol with carbon capture 
& Biomass-based methanol production with capture of process CO$_2$. 
& Biofuels \\

Biogas to methanol 
& Conversion of biogas-derived carbon into methanol without carbon capture. 
& Biofuels \\

Biogas to methanol with carbon capture 
& Biogas-based methanol production combined with capture of CO$_2$. 
& Biofuels \\

E-biomethanol 
& Methanol produced by adding renewable hydrogen to the biomass-to-methanol process to increase carbon utilisation efficiency. 
& E-biofuels \\

E-biomass-to-liquid, 
& Liquid fuels produced by integrating renewable hydrogen directly into biomass-to-liquid conversion processes. 
& E-biofuels \\

E-biogas-methanol 
& Methanol produced from biogas-derived carbon with renewable hydrogen addition. 
& E-biofuels \\

Fischer--Tropsch
& Synthetic liquid fuels produced from renewable hydrogen and captured CO$_2$. 
& E-fuels \\

Methanolisation 
& Synthetic methanol produced from renewable hydrogen and captured CO$_2$. 
& E-fuels \\

Blue methanol 
& Methanol produced from fossil feedstocks with carbon capture and storage. 
& Fossil fuels \\

Grey methanol 
& Methanol produced from fossil feedstocks without carbon capture. 
& Fossil fuels \\

Oil refining 
& Refining of crude oil into conventional liquid fuels. 
& Fossil fuels \\
\bottomrule
\end{tabular}

\caption{Definitions and classification of liquid fuel technologies considered in this study. Biofuels rely exclusively on biogenic carbon, e-biofuels combine biogenic carbon with renewable hydrogen, e-fuels are produced entirely from renewable hydrogen and captured CO$_2$, while fossil fuels are derived from fossil feedstocks with or without carbon capture.}
\label{tab:S1_liquid_fuel_definitions}
\end{table}

\end{document}